\documentclass[11pt,a4paper]{article}
\usepackage{authblk}
\usepackage{lmodern}
\usepackage{jheppub}
\usepackage{amsmath}
\usepackage{mathtools}
\usepackage{tikz}
\usepackage{empheq}
\usepackage{enumitem}
\usepackage{graphics}
\allowdisplaybreaks
\usepackage{wrapfig}
\usepackage{caption}
\usepackage{natbib}
\usepackage{xcolor}
\usepackage{framed}
\usepackage{array}
\usepackage{dsfont}
\definecolor{shadecolor}{gray}{0.925}
\usepackage[cbgreek]{textgreek}

\def\sideremark#1{\ifvmode\leavevmode\fi\vadjust{\vbox to0pt{\vss
 \hbox to 0pt{\hskip\hsize\hskip1em
 \vbox{\hsize3cm\tiny\raggedright\pretolerance10000
 \noindent #1\hfill}\hss}\vbox to8pt{\vfil}\vss}}}%

\newcommand{\bi}{\begin{itemize}}
\newcommand{\ei}{\end{itemize}}
\newcommand{\bea}{\begin{align}}
\newcommand{\eea}{\end{align}}
\newcommand{\be}{\begin{equation}}
\newcommand{\ee}{\end{equation}}

\newcommand{\tcb}{\textcolor{blue}}

\makeatletter
\renewcommand*\env@matrix[1][\arraystretch]{%
  \edef\arraystretch{#1}%
  \hskip -\arraycolsep
  \let\@ifnextchar\new@ifnextchar
  \array{*\c@MaxMatrixCols c}}
\makeatother
\author[\ensuremath{c},\ensuremath{d}]{Charlotte SLEIGHT}
\author[\ensuremath{a},\ensuremath{b},\ensuremath{c}]{\quad Massimo TARONNA}

\affiliation[\ensuremath{a}]{Dipartimento di Fisica ``Ettore Pancini'', Universit\`a degli Studi di Napoli Federico II, \\Monte S. Angelo, Via Cintia, 80126 Napoli, Italy}

\affiliation[\ensuremath{b}]{Scuola Superiore Meridionale, Universit\`a degli Studi di Napoli Federico II,\\ Largo San Marcellino 10, 80138 Napoli, Italy}

\affiliation[\ensuremath{c}]{INFN, Sezione di Napoli, Monte S. Angelo, Via Cintia, 80126 Napoli, Italy}

\affiliation[\ensuremath{d}]{Centre for Particle Theory and Department of Mathematical Sciences, \\ Durham University, Durham, DH1 3LE, U.K.}

\emailAdd{charlotte.sleight@durham.ac.uk, massimo.taronna@unina.it}

\title{\centering \huge Celestial Holography Revisited}

\abstract{We revisit the prescription commonly used to define holographic correlators on the celestial sphere of
Minkowski space as an integral transform of flat space scattering amplitudes, known as celestial amplitudes. We propose a resolution to a discrepancy noted in the computation of celestial amplitudes, which arises from the regularisation and the commutation of a divergent integral in the definition of conformal primary wave functions. Motivated by this, we propose a novel, off-shell, prescription for holographic correlators on the celestial sphere which we refer to as \emph{celestial correlators}. The latter are defined by the Mellin transform of bulk
time-ordered correlators with respect to the radial direction in the hyperbolic slicing of Minkowski space,
which are then extrapolated to the celestial sphere along the hyperbolic directions. This prescription is
analogous to the extrapolate definition of holographic correlators in AdS/CFT and, like in AdS, is centered
on (off-shell) correlation functions as opposed to (on-shell) S-matrix elements. We show that celestial
correlators defined in this new way are manifestly recast in terms of corresponding Witten diagrams in
Euclidean anti–de Sitter space in perturbation theory. We discuss the possibility of using this definition
of celestial correlators in terms of bulk correlation functions to explore the non-perturbative properties of
celestial correlators dual to conformal field theories in Minkowski space. We furthermore show that celestial amplitudes can also be defined by a similar extrapolation of S-matrices in position space via a Mellin transformation in the radial direction. This provides the proper regularisation of conformal primary wave functions, which is inherited from the corresponding Wightman functions.}

\begin{document}

\begin{flushright}    
\texttt{}
\end{flushright}

\maketitle

\newpage

\section{Introduction}\label{sec::Intro}

\begin{figure}[t]
    \centering
    \includegraphics[width=0.65\textwidth]{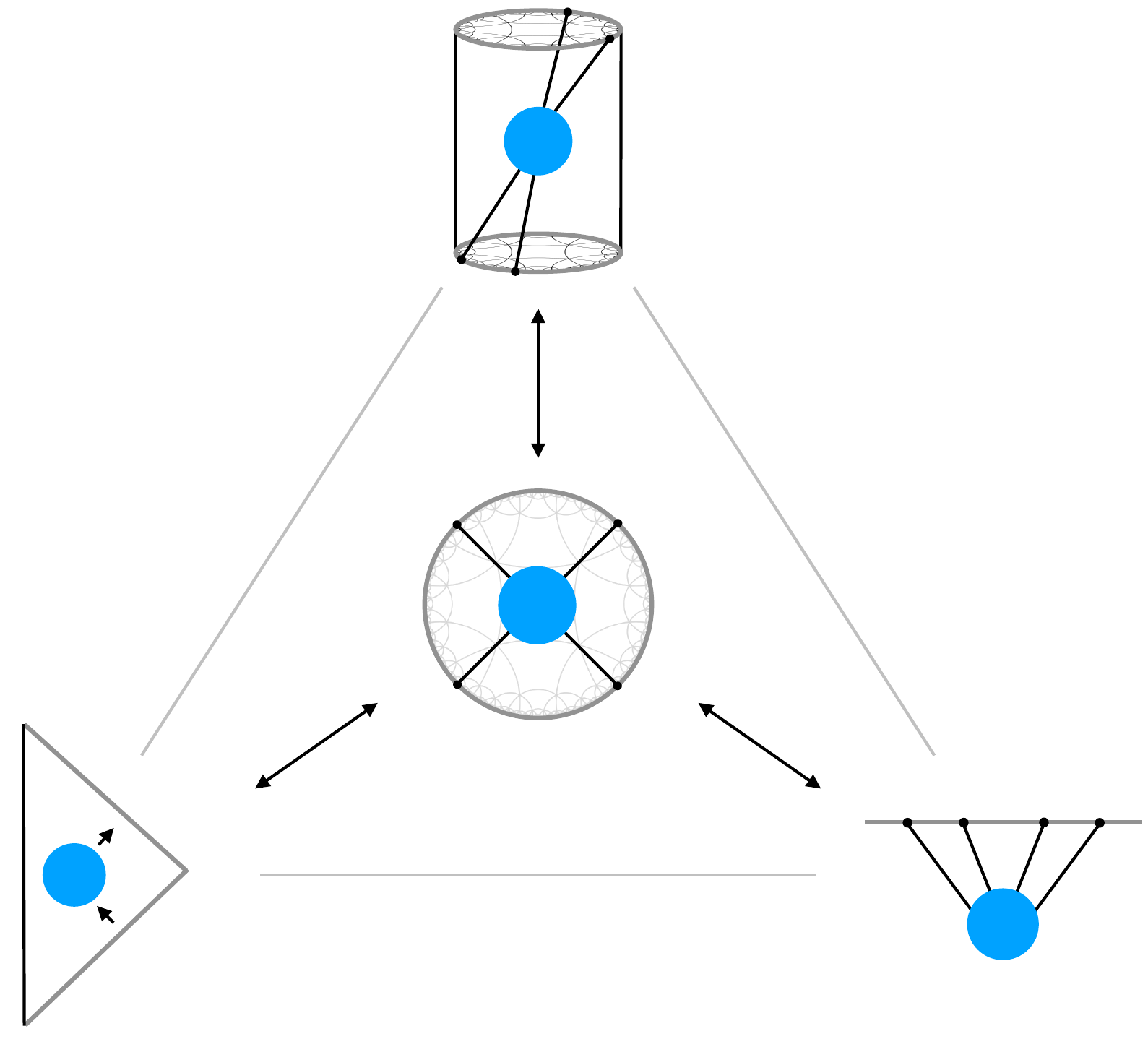}
    \caption{The Holographic Triangle: A cartoon representation of how perturbative observables at the conformal boundary in different maximally symmetric spaces can be recast in terms of boundary correlators in EAdS (centre). Top: $\Lambda <0$, bottom-right: $\Lambda >0$ and bottom-left: $\Lambda=0$.}
    \label{fig::holotri}
\end{figure}

The holographic principle is a very powerful idea which provides a framework to study quantum gravity observables living on the conformal boundary of space-time. This has been most successfully applied in the context of anti-de Sitter (AdS) space, where the AdS/CFT correspondence \cite{Maldacena:1997re,Gubser:1998bc,Witten:1998qj} conjectures that quantum gravity observables on the boundary of AdS$_{d+1}$ space are equivalent to correlation functions of a (non-gravitational) Conformal Field Theory (CFT) in $d$-dimensional Minkowski space $\mathbb{M}^d$.

\vskip 4pt
A key feature of AdS space is that its boundary lies at spatial infinity. The dual boundary theory is therefore a standard quantum mechanical system with a standard notion of locality and of time. This poses a key hurdle along the way to extend our understanding of holographic quantum gravity observables beyond the relative security of the AdS/CFT correspondence. Recent years have seen significant progress in the context of anti-de Sitter's maximally symmetric (and more realistic) cousins Minkowski space and de Sitter (dS) space, which have come to be known as Celestial Holography \cite{Raclariu:2021zjz,Pasterski:2021rjz,McLoughlin:2022ljp,Pasterski:2021raf} and the Cosmological Bootstrap \cite{Baumann:2022jpr,Benincasa:2022gtd}, respectively. In contrast to AdS space, the conformal boundaries of Minkowski and dS space lie at null and past/future infinity, respectively, which obscures how the corresponding boundary correlation functions encode consistent bulk physics. Much of this progress has been driven by the discovery of connections with more familiar flat space S-matrices. In the context of celestial holography, correlators on the celestial sphere have been defined as a conformal change of basis of flat space scattering amplitudes \cite{deBoer:2003vf,Cheung:2016iub,Pasterski:2016qvg,Pasterski:2017kqt}, known as \emph{celestial amplitudes}. 

\vskip 4pt
Another promising approach has been to draw connections with the well known perturbative computation of boundary correlators in AdS/CFT via Witten diagrams \cite{Cheung:2016iub,Lam:2017ofc,Sleight:2019hfp,Sleight:2020obc,DiPietro:2021sjt,Sleight:2021plv,Casali:2022fro,deGioia:2022fcn,Iacobacci:2022yjo}. It has been shown \cite{Sleight:2020obc,Sleight:2021plv,Iacobacci:2022yjo} that both dS boundary correlators in the Bunch-Davies (Euclidean) vacuum and celestial amplitudes can be perturbatively recast as Witten diagrams in Euclidean AdS (EAdS). This approach, which places correlators on the boundary of EAdS at the centre, has been dubbed ``the holographic triangle" (see figure \ref{fig::holotri}). This is the idea that reformulating boundary correlators in AdS, Minkowski and dS as boundary correlators in EAdS could provide a way to place holography for $\Lambda<0$, $\Lambda=0$ and $\Lambda>0$ (i.e. for \emph{all} $\Lambda$s) on a similar footing. How consistency criteria such as unitarity and causality are encoded in such EAdS boundary correlators depends on where the original theory is defined in the bulk (i.e Lorentzian AdS, Minkowski or de Sitter space).

\vskip 4pt
In the context of celestial holography, it was recently noted \cite{Iacobacci:2022yjo} that there seems to be a discrepancy between these two approaches to celestial amplitudes. In one case, celestial amplitudes are obtained from the corresponding S-matrix via an integral transform, while in the other celestial amplitudes are treated analogously to Witten diagrams with the conformal primary wave functions treated like bulk-to-boundary propagators. This discrepancy appears to arise from the fact that the relationship between celestial amplitudes and momentum space scattering amplitudes commutes divergent integrals forming part of the definition of conformal primary wavefunctions. In this work we re-visit this discrepancy and point out that the regularisation introduced to define conformal primary wave functions in \cite{Pasterski:2016qvg,Pasterski:2017kqt} alters the nature of the observable that is being considered on the celestial sphere; the regularisation appears to implicitly select a time-ordering and takes the external legs off-shell. Celestial amplitudes computed using the regularised expression \cite{Pasterski:2016qvg,Pasterski:2017kqt} for conformal primary wave functions are therefore (anti-)time-ordered propagators with one leg extrapolated to the celestial sphere.

\vskip 4pt
In light of this, we propose a novel, off-shell, extrapolate definition of correlators on the celestial sphere, in which the conformal primary wave functions are replaced with analytic functions obtained by extrapolating one point of the Minkowski Feynman propagator $G_T$ to the celestial sphere. In practise, this is achieved by taking a Mellin transform with respect to the radial direction in the hyperbolic slicing \eqref{hypslice} of Minkowski space and the boundary limit in the hyperbolic directions:
\begin{equation}\label{celestialbubointro}
    G^{\text{flat}}_{\Delta}\left(X,Q\right) =\lim_{\hat{Y}\to Q} \int^\infty_0 \frac{dt}{t} t^\Delta\,G_T\left(X,t\hat{Y}\right).
\end{equation}
 We refer to the functions $G^{\text{flat}}_{\Delta}$ as \emph{celestial bulk-to-boundary propagators}. According to this definition, correlators on the celestial sphere are given by a Mellin transform of bulk time-ordered correlators in Minkowski space that are extrapolated to the boundary:
\begin{align}\label{celestialcorrelationnewdef0}
    \left\langle\mathcal{O}_{\Delta_1}(Q_1)\ldots \mathcal{O}_{\Delta_n}(Q_n)\right\rangle =\prod_i \lim_{{\hat Y}_i\to Q_i}\,\int^\infty_0 \frac{dt_i}{t_i}\,t_i^{\Delta_i}\left\langle\phi_1(t_1\hat{Y}_1)\ldots \phi_n(t_n\hat{Y}_n)\right\rangle.
\end{align}
We refer to such off-shell observables on the celestial sphere as \emph{celestial correlators}, to distinguish them from (on-shell) \emph{celestial amplitudes}. This definition naturally extends to celestial holography the extrapolate definition of boundary correlation functions commonly employed in (A)dS/CFT. 

\vskip 4pt
The definition \eqref{celestialcorrelationnewdef0} of celestial correlators places correlation functions at the centre of celestial holography. This has the advantage that it can be applied to theories in Minkowski space for which the S-matrix is not defined. An important example of such theories are Conformal Field Theories, which in Minkowski space are defined non-perturbatively at the level of correlation functions by Conformal Symmetry, Unitarity and a consistent operator product expansion. The definition \eqref{celestialcorrelationnewdef0} then opens up the possibility to study the properties of celestial correlation functions from their relatively well understood Minkowski CFT counterparts.

\vskip 4pt
Coming back to the holographic triangle (figure \ref{fig::holotri}), celestial correlators defined according to \eqref{celestialcorrelationnewdef0} can also be perturbatively re-cast as corresponding Witten diagrams in EAdS$_{d+1}$ along the same lines as \cite{Iacobacci:2022yjo} for celestial amplitudes. In the case of celestial correlators, the relationship to EAdS Witten diagrams is more manifest. In the hyperbolic slicing of Minkowski space, the hyperbolic dependence of the celestial bulk-to-boundary propagators \eqref{celestialbubointro} is given by the corresponding (time-ordered) bulk-to-boundary propagator in (EA)dS$_{d+1}$ and a radial dependence given by the kernel of the Kontorovich-Lebedev transform. The result then follows from the de Sitter case \cite{Sleight:2020obc,Sleight:2021plv}. In this work this is demonstrated for contact diagram contributions to celestial correlators, which can be expressed in terms of contact Witten diagrams in EAdS. The proof for any given perturbative contribution to celestial correlators \eqref{celestialcorrelationnewdef0} is given in the follow-up work \cite{Iacobacci:2024nhw}.

\vskip 4pt
Finally, we note that the extrapolation \eqref{celestialbubointro} of bulk points to the celestial sphere via radial Mellin transform can also be used to define conformal primary wave functions $\phi^\pm_{\Delta}\left(X;Q\right)$ as the extrapolation of corresponding two-point Wightman functions $\mathcal{W}^\pm(X,Y)$:
\begin{equation}\label{schwbubo0}
    \phi^\pm_{\Delta}\left(X;Q_\pm\right)=\lim_{\hat{Y}\to Q_\pm}\int_{0}^\infty \frac{dt}{t} t^{\Delta}\,\mathcal{W}^\pm(X,t\hat{Y})\,.
\end{equation}
Celestial amplitudes can then be defined as the extrapolation \eqref{celestialcorrelationnewdef0} of position space scattering amplitudes:
\begin{equation}\label{CAradial}
    {\cal A}_{\Delta_1, \ldots,\Delta_n}\left(Q^\pm_1, \ldots , Q^\pm_n\right) = \prod_i \lim_{{\hat Y}_i\to Q^\pm_i}\,\int^\infty_0 \frac{dt_i}{t_i}\,t_i^{\Delta_i}\,{\cal A}\left(t_1{\hat Y}_1, \ldots , t_n{\hat Y}_n\right),
\end{equation}
where e.g. for massive fields
\begin{equation}
    {\cal A}\left(Y_1, \ldots , Y_n\right) = \prod^n_{i=1} \int_{P^2_i+m^2_i=0,\, P^0_i>0} \frac{d^{d+2}P_i} {\left(2\pi\right)^{d+2}} e^{\pm i P_i \cdot Y_i} {\cal A}\left(\pm P_1, \ldots , \pm P_n\right). 
\end{equation}
The definition \eqref{CAradial} of celestial amplitudes clarifies that the correct regularisation of conformal primaries is inherited from the corresponding Wightman function. This is detailed in section \ref{sec::CPWFrevisited}.

\newpage

\section{Hyperbolic slicing of Minkowski space}
\label{sec::HSMINK}

\begin{figure}[htb]
    \centering
    \includegraphics[width=0.6\textwidth]{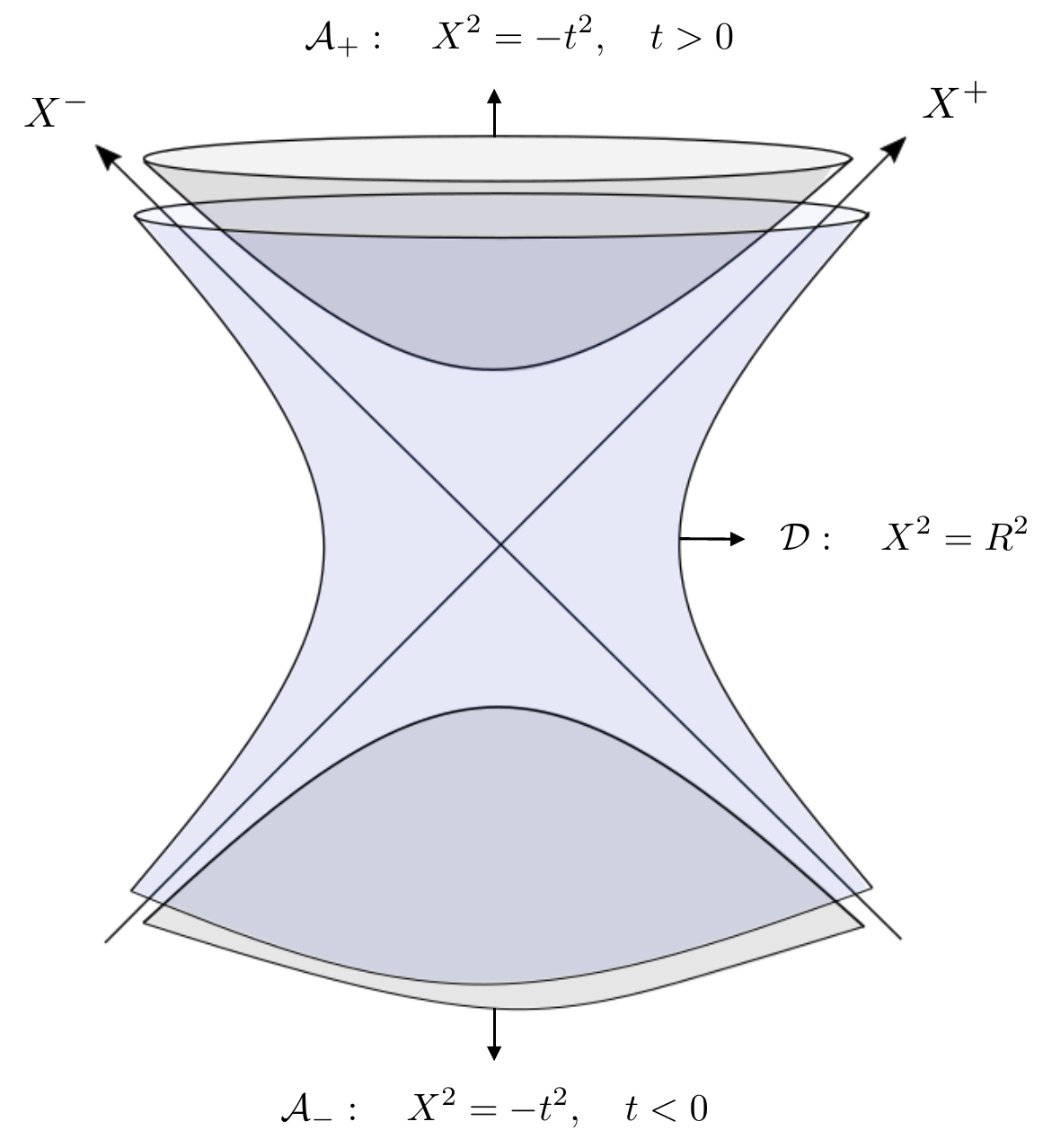}
    \caption{Hyperbolic foliation of Minkowski space with light-cone coordinates $X^\pm$.}
    \label{fig::HMink}
\end{figure}

In this work we consider $\left(d+2\right)$-dimensional Minkowski space $\mathbb{M}^{d+2}$ with Cartesian coordinates $X^M$, $M=0,\ldots, d+1$ and metric
\begin{equation}
    ds^2=-\left(dX^0\right)^2+\left(dX^1\right)^2+\ldots+\left(dX^{d+1}\right)^2. 
\end{equation} 
Following \cite{deBoer:2003vf}, a radial foliation of $\mathbb{M}^{d+2}$ is naturally achieved by considering the following three regions:\begin{subequations}\label{hypslice}
\begin{align}
    \mathcal{D}:&\quad X^2>0\,,\\
    \mathcal{A}_+:&\quad X^2<0\,,\qquad X^0>0,\\
    \mathcal{A}_-:&\quad X^2<0\,,\qquad X^0<0.
\end{align}
\end{subequations}
Each region of these regions can be foliated with surfaces of constant curvature reflecting the $SO\left(d+1,1\right)$ symmetry. See figure \ref{fig::HMink}. In regions $\mathcal{A}_\pm$ these are $\left(d+1\right)$-dimensional Euclidean anti-de Sitter spaces (EAdS$_{d+1}$) with constant radius $t$:
\begin{equation}
    X^2=-t^2. \label{EAdSt}
\end{equation}
A natural set of coordinates for this foliation of ${\cal A}_\pm$ which is particularly well-suited for holography is given by Poincar\'e coordinates
\begin{subequations}\label{EAdScoords}
\begin{align}
    \mathcal{A}_+:&\qquad X^M=+\frac{t}{z}\left(\frac{1+z^2+\vec{x}^2}{2},\frac{1-z^2-\vec{x}^2}{2},\vec{x}\right),& t&>0,\quad z>0,\\
    \mathcal{A}_-:&\qquad X^M=+\frac{t}{z}\left(\frac{1+z^2+\vec{x}^2}{2},\frac{1-z^2-\vec{x}^2}{2},\vec{x}\right),& t&<0,\quad z>0.
\end{align}
\end{subequations}
For region ${\cal D}$ instead, the foliating surfaces are $\left(d+1\right)$-dimensional de Sitter space-times with radius $R$:
\begin{equation}
    X^2=R^2. \label{dSr}
\end{equation}
It is convenient to cover each dS space with two Poincar\'e patches:
\begin{subequations}\label{dScoords}
\begin{align}
    \mathcal{D}_+:&\qquad X^M=\frac{R}{(-\eta)}\left(\frac{1-\eta^2+\vec{x}^2}{2},\frac{1+\eta^2-\vec{x}^2}{2},\vec{x}\right),& R&>0,\quad \eta<0\\
    \mathcal{D}_-:&\qquad X^M=-\frac{R}{\eta}\left(\frac{1-\eta^2+\vec{x}^2}{2},\frac{1+\eta^2-\vec{x}^2}{2},\vec{x}\right),& R&>0,\quad \eta>0 
\end{align}
\end{subequations}
The two regions $\mathcal{D}_\pm$ of ${\cal D}$ correspond to expanding and contracting patches of the dS hypersurface \eqref{dSr} respectively:
\begin{subequations}
\begin{align}
    \mathcal{D}_+:&\quad X^2>0\,,\qquad X^+>0,\\
    \mathcal{D}_-:&\quad X^2>0\,,\qquad X^+<0,
\end{align}
\end{subequations}
where $X^+$ is the light cone coordinate $X^+=X^{0}+X^{1}$. 

\vskip 4pt
\paragraph{Conformal Boundary.} $\mathbb{M}^{d+2}$ has a conformal boundary at past and future null infinity, which is identified with the projective cone of light rays via
\begin{equation}\label{PrLC}
    Q^2=0, \qquad Q \equiv \lambda Q, \qquad \lambda \in \mathbb{R}_+.
\end{equation}
The past and future conformal boundaries are both $d$-dimensional spheres, which we shall denote by $S^-_d$ and $S^+_d$ respectively. To see this one introduces new projective coordinates
\begin{align}
    \xi_1 = Q^1/Q^0, \quad \xi_2 = Q^2/Q^0, \quad \ldots \quad, \quad \xi_{d+1} = Q^{d+1}/Q^0,
\end{align}
so that 
\begin{equation}
    \xi^2_1+\ldots+\xi^2_{d+1}-1=0.
\end{equation}
For the sphere $S^-_d$ we take $Q^0<0$ and for $S^+_d$ we take $Q^0>0$. These conformal boundaries of $\mathbb{M}^{d+2}$ are also conformal boundaries of each of its hyperbolic slices \eqref{EAdSt} and \eqref{dSr}, which manifest from the fact that the slices asymptote to the lightcone \eqref{PrLC}. The region ${\cal A}_+ \left({\cal A}_-\right)$ is foliated by the upper (lower) sheet of the hyperboloids \eqref{EAdSt} and have conformal boundary $S^+_d \left(S^-_d\right)$ at spatial infinity. In the parameterisations \eqref{EAdScoords} the conformal boundary is reached by sending $z \to 0$ with boundary coordinates:
\begin{subequations}
\begin{align}
    Q_+&\sim \frac{t}{z}\, \left(\frac{1+x^2}2,\frac{1-x^2}2,\vec{x}\right),&  &\text{with}&  t>0,\\
    Q_-&\sim \frac{t}{z}\,\left(\frac{1+x^2}2,\frac{1-x^2}2,\vec{x}\right),&  &\text{with}& t<0.
\end{align}
\end{subequations}
In region ${\cal D}$ the foliating surfaces \eqref{dSr} are de Sitter space-times, which each have a conformal boundary at future infinity $\left(S^+_d\right)$ and past infinity $\left(S^-_d\right)$. In the parameterisation \eqref{dScoords} these are obtained in the limit $\eta \to 0$ with boundary coordinates:
\begin{subequations}
\begin{align}
    Q_+&\sim \frac{R}{(-\eta)}\left(\frac{1+x^2}2,\frac{1-x^2}2,\vec{x}\right),&  &\text{with}&  \eta<0, \\Q_- &\sim -\frac{R}{\eta}\left(\frac{1+x^2}2,\frac{1-x^2}2,\vec{x}\right), &  &\text{with}&  \eta>0.
\end{align}
\end{subequations}
Note that by working in complexified space one can analytically continue $Q_+$ to $Q_-$ by considering:
\begin{align}
    Q_\theta=e^{i\theta}\left(\frac{1+x^2}2,\frac{1-x^2}2,\vec{x}\right)\,.
\end{align}
Working in a complexified space-time will be useful when considering the analytic continuation in the next sections of this work.

\section{Conformal Primary Wave Functions Revisited}
\label{sec::CPWFrevisited}

In \cite{Pasterski:2016qvg,Pasterski:2017kqt} celestial amplitudes were defined as an integral transform of flat space scattering amplitudes by introducing the following conformal primary wave functions:
\begin{equation}\label{ConfMode0}
    \phi^\pm_{\Delta}\left(X;Q_\pm\right)=\int_{P^2+m^2=0,\, P^0>0} \frac{d^{d+2}P}{\left(2\pi\right)^{d+2}} \frac{\pi \Gamma(\Delta)}{(\pm iP\cdot Q_\pm)^\Delta}\,e^{\pm i P\cdot X},
\end{equation}
where $Q_\pm \in S^\pm_d$. This expression provides the change of basis from plane waves to conformal primary wave functions, which transform like conformal primaries under Lorentz transformations.

\vskip 4pt
It will be instructive to note that the conformal primary wave functions \eqref{ConfMode0} can in fact be obtained from the two-point Wightman function,
\begin{equation}\label{Wightman}
    \mathcal{W}^\pm(X,Y) = \int \frac{d^{d+2}P}{\left(2\pi \right)^{d+2}}\,\theta(P^0)\pi \delta(P^2+m^2)\,e^{\pm iP\cdot(X-Y)}.
\end{equation}
This is achieved by sending one of the two bulk points, say $Y$, to the celestial sphere. To this end, one considers the hyperbolic foliation $Y = t {\hat Y}$ and takes a Mellin transform with respect to the radial direction $t$ and extrapolating the remaining hyperbolic directions $\hat{Y}$ to the conformal boundary:\footnote{This can be seen by simply noting that: 
\begin{equation}\label{schwbubo}
    \int_{0}^\infty \frac{dt}{t} t^{\Delta}\, e^{\mp i t P \cdot Q} = \frac{\Gamma\left(\Delta\right)}{\left(\pm i P \cdot Q + \epsilon \right)^{\Delta}}.
\end{equation}}
\begin{equation}\label{extrapW}
    \phi^\pm_{\Delta}\left(X;Q_\pm\right)=\lim_{\hat{Y}\to Q_\pm}\int_{0}^\infty \frac{dt}{t} t^{\Delta}\,\mathcal{W}^\pm(X,t\hat{Y})\,.
\end{equation}
In this work we shall refer to such a Mellin transform in the radial direction of $\mathbb{M}^{d+2}$ as a \emph{radial Mellin transform}.

\vskip 4pt
The defining integral \eqref{ConfMode0} is divergent. Introducing a regulator the following expression was obtained in \cite{Pasterski:2016qvg,Pasterski:2017kqt} in terms of the Bessel-$K$ function\footnote{Note that we adopt a different normalisation to \cite{Pasterski:2016qvg,Pasterski:2017kqt} for the conformal primary wave function. In the present paper it is instead fixed by the Wightman function \eqref{Wightman} via \eqref{extrapW}.} 
\begin{align}\label{regdefcpw}
    \phi^\pm(X,Q)=\frac{1}{2\pi^{\frac{d+2}{2}}}\left(\frac{m}{2}\right)^{\frac{d}{2}-\Delta}\frac{\Gamma(\Delta)}{(-2X\cdot Q\mp i\epsilon)^\Delta} (\sqrt{X^2\mp i\epsilon})^{\Delta-\frac{d}{2}} {K}_{\Delta-\tfrac{d}2}(m\sqrt{X^2\mp i\epsilon})\,.
\end{align}
This expression however appears to give rise to an inconsistency in the definition of the conformal primary wave function, which can be seen by comparing its Fourier transform with that of the original definition \eqref{ConfMode0}. Instead of recovering \eqref{ConfMode0}, one obtains an off-shell  (anti-)time ordered two-point function (see appendix \ref{A::CPWFs}): 
\begin{align}\label{fouriercpwps}
\tilde{\phi}^\pm_{\Delta}(P,Q) &= \int \frac{d^{d+2} P}{\left(2\pi \right)^{d+2}} \phi^\pm_{\Delta}(X,Q) e^{+i P \cdot X}\\
   &=\pm i\frac{\Gamma\left(\Delta\right)}{\left(i P \cdot Q + \epsilon\right)^{\Delta}}\frac{1}{P^2+m^2\pm i \epsilon},
\end{align}
which is the Fourier transform of the (anti)-time-ordered propagator $G_{({\bar T})T}$ in Minkowski space with one bulk point extrapolated to the celestial sphere. In other words, instead of \eqref{extrapW} one finds that${}^{\tcb{1}}$
\begin{align}
    \int d^{d+2} X \tilde{\phi}^+_{\Delta}(P,Q) e^{-i P \cdot X} &= \lim_{\hat{Y}\to Q} \int^\infty_0 \frac{dt}{t} t^\Delta\,G_{\bar T}\left(X,t\hat{Y}\right),\\
  \int d^{d+2} X \tilde{\phi}^-_{\Delta}(P,Q) e^{-i P \cdot X} &=  \lim_{\hat{Y}\to Q} \int^\infty_0 \frac{dt}{t} t^\Delta\,G_T\left(X,t\hat{Y}\right),
\end{align}
where 
\begin{align}
    G_T(X,Y)&=\int \frac{d^{d+2}P}{(2\pi)^{d+2}}\frac{-i}{P^2+m^2-i\epsilon}\,e^{iP\cdot(X-Y)},\\
    G_{\bar T}(X,Y)&=\int \frac{d^{d+2}P}{(2\pi)^{d+2}}\frac{+i}{P^2+m^2+i\epsilon}\,e^{iP\cdot(X-Y)}.
\end{align}
We see that the $i\epsilon$ prescription introduced in \eqref{regdefcpw} to define the conformal primaries is equivalent to selecting a time-ordering for the 2pt function. It is the time-ordered Feynman propagator in the case of $\phi^-$ and the anti-time-ordered Feynman propagator in the case of $\phi^+$. This can be seen more clearly by considering the following re-writing of the Wightman functions:
\begin{align}
    W^+\left(X;Y\right) &= \theta \left(X^0 - Y^0\right) G_{T}\left(X;Y\right)+\theta \left(Y^0 - X^0\right) G_{{\bar T}}\left(X;Y\right),\\
    W^-\left(X;Y\right) &= \theta \left(X^0 - Y^0\right) G_{{\bar T}}\left(X;Y\right)+\theta \left(Y^0 - X^0\right) G_{T}\left(X;Y\right),
\end{align}
where, by comparing with \eqref{fouriercpwps}, we can see that the conformal primary wave function \eqref{regdefcpw} corresponds to the ordering $X^0-Y^0 <0$. We conclude that celestial amplitudes defined by the regularised conformal primary wave functions \eqref{regdefcpw} are in fact the extrapolation to the celestial sphere of off-shell, \emph{out-of-time-ordered}, correlation functions in $\mathbb{M}^{d+2}$. 

\vskip 4pt
 In the light of this observation, in this work we propose an off-shell definition of correlators on the celestial sphere that manifestly respects the time-ordering of the bulk points in flat space. Such a definition would be given by conformal modes that are instead the extrapolation via radial Mellin transform of the \emph{Feynman propagator} $G_T$: 
\begin{equation}\label{celestialbubo}
    G^{\text{flat}}_{\Delta}\left(X,Q\right) =\lim_{\hat{Y}\to Q} \int^\infty_0 \frac{dt}{t} t^\Delta\,G_T\left(X,t\hat{Y}\right),
\end{equation}
where $Q$ resides on either $S^+_d$ or $S^-_d$. We refer to \eqref{celestialbubo} as the \emph{celestial bulk-to-boundary propagator}. Correlators on the celestial sphere defined with respect to the celestial bulk-to-boundary propagator \eqref{celestialbubo} are then a radial Mellin transform of \emph{time-ordered} correlation functions in $\mathbb{M}^{d+2}$, which are extrapolated to the conformal boundary in the hyperbolic directions. We refer to such correlators as \emph{celestial correlators} and they are defined by the formula:
\begin{align}
    \left\langle\mathcal{O}_{\Delta_1}(Q_1)\ldots \mathcal{O}_{\Delta_n}(Q_n)\right\rangle =\prod_i \lim_{{\hat Y}_i\to Q_i}\,\int^\infty_0 \frac{dt_i}{t_i}\,t_i^{\Delta_i}\left\langle\phi_1(t_1\hat{Y}_1)\ldots \phi_n(t_n\hat{Y}_n)\right\rangle.
\end{align}
The Feynman rules for such celestial correlation functions are then inherited from those of time-ordered correlation functions with celestial bulk-to-boundary propagators \eqref{celestialbubo} on the external legs.

\vskip 4pt
In appendix \ref{appendix::CBUBO} we show that the celestial bulk-to-boundary propagator can be expressed explicitly in the following form
\begin{align}\label{explcelesbubo}
    G^{\text{flat}}_{\Delta}(X,Q)
    &=c_\Delta^{\text{dS-AdS}} G^{\text{AdS}}_\Delta(\hat{X}_\epsilon,Q)\,\mathcal{K}^{(m)}_{i\left(\frac{d}{2}-\Delta\right)}\left(\sqrt{X^2+i\epsilon}\right)\,,
\end{align}
in terms of the the kernel $\mathcal{K}^{(m)}_{i\left(\frac{d}{2}-\Delta_i\right)}$ of the Kontorovich-Lebedev transform (see Appendix \ref{appendix::KL-transform}) and the corresponding (analytically continued) EAdS$_{d+1}$ bulk-to-boundary propagator:
\begin{align}
    G^{\text{AdS}}_\Delta(\hat{X}_\epsilon,Q)=
   C^{\text{AdS}}_\Delta \frac{\left(\sqrt{X^2+i\epsilon}\right)^{\Delta}}{(-2 X\cdot Q+i\epsilon)^\Delta}\,,
\end{align}
with 
\begin{align}\label{AdS2pt}
    C^{\text{AdS}}_\Delta&=\frac{\Gamma (\Delta )}{2\pi^{d/2} \Gamma \left(\Delta-\frac{d}{2} +1\right)}\,.
\end{align}
The coefficient 
\begin{equation}
     c_\Delta^{\text{dS-AdS}}=\frac{C^{\text{dS}}_{\Delta}}{C^{\text{AdS}}_{\Delta}}=\frac12\,\csc(\pi(\tfrac{d}2-\Delta)),
\end{equation}
accounts for the difference in two-point function normalisation between AdS \eqref{AdS2pt} and dS \cite{Sleight:2021plv}. The explicit expression \eqref{explcelesbubo} makes manifest that celestial correlators can (at least perturbatively) be re-cast as Witten diagrams in EAdS$_{d+1}$ appropriately analytically continued along the complexified null cone. This is demonstrated in the next section for contact diagrams, where the proof to all orders in perturbation theory is given in the follow-up paper \cite{Iacobacci:2024nhw}.

\vskip 4pt
Before concluding this section, let us note that the radial Mellin transform can also be used to define celestial amplitudes. One first obtains the position-space scattering amplitude, which e.g. for massive particles reads
\begin{equation}
    {\cal A}\left(X_1, \ldots , X_n\right) = \prod^n_{i=1} \int_{P^2_i+m^2_i=0,\, P^0_i>0} \frac{d^{d+2}P_i} {\left(2\pi\right)^{d+2}} e^{\pm i P_i \cdot X_i} {\cal A}\left(\pm P_1, \ldots , \pm P_n\right). 
\end{equation}
Celestial amplitudes can then be obtained by extrapolating the bulk points $X_i$ to the celestial sphere according to \eqref{extrapW}:
\begin{equation}\label{MTcelestialampl}
    {\cal A}_{\Delta_1, \ldots,\Delta_n}\left(Q^\pm_1, \ldots , Q^\pm_n\right) = \prod_i \lim_{{\hat X}_i\to Q^\pm_i}\,\int^\infty_0 \frac{dt_i}{t_i}\,t_i^{\Delta_i}\,{\cal A}\left(t_1{\hat X}_1, \ldots , t_n{\hat X}_n\right).
\end{equation}
Using \eqref{schwbubo} one then recovers the definition of celestial amplitudes given in \cite{Pasterski:2016qvg,Pasterski:2017kqt}:
\begin{multline}\label{celestialampl}
    {\cal A}_{\Delta_1, \ldots,\Delta_n}\left(Q_1, \ldots , Q_n\right) = \prod^n_{i=1} \int_{P^2_i+m^2_i=0,\, P^0_i>0} \frac{d^{d+2}P_i} {\left(2\pi\right)^{d+2}}\frac{\Gamma\left(\Delta_i\right)}{\left(\mp i P_i \cdot Q^\pm_i + \epsilon\right)^{\Delta_i}}\\ \times {\cal A}\left(\pm P_1, \ldots , \pm P_n\right).
\end{multline}
This clarifies that the correct regularisation of the conformal primary wave functions is provided by the corresponding Wightman function via \eqref{extrapW}.

\vskip 4pt
The expressions for celestial amplitudes obtained in \cite{Iacobacci:2022yjo} instead worked directly with the regularised definition \eqref{regdefcpw} of conformal primary wave functions. This explains their discrepancy with the expressions for the same celestial amplitudes obtained in \cite{Pasterski:2016qvg}, which were obtained by employing directly the definition \eqref{celestialampl} instead.

\section{Celestial Correlators}
\label{sec::ccrev}

In this section we explore in more detail celestial correlators defined according to \eqref{celestialcorrelationnewdef0}. We give some examples in perturbation theory for massive scalar field theories and show that they can be recast in terms of corresponding Witten diagrams in Euclidean AdS suitably analytically continued along the complex light-cone.

\subsection{Two-point function normalisation}
\label{subsec::2pt}

Let us first determine the normalisation of the celestial two-point function.\footnote{For simplicity, in this section we do not consider the case that the two operators are shadow of one another i.e. $\Delta_1 +\Delta_2= d$, which is equivalent to consider $\Delta_i=\frac{d}{2}+i\nu_i$ with $\nu_i>0$. This can non-the-less be obtained explicitly in a similar fashion.} This is obtained from the Feynman propagator by extrapolating both bulk points to the celestial sphere (appendix \ref{sec::celestial2pt}):
\begin{align}
 \langle \mathcal{O}_{\Delta_1}\left(Q_1\right) \mathcal{O}_{\Delta_2}\left(Q_2\right) \rangle &= \lim_{{\hat X}_i \to Q_i}\int_0^\infty \frac{dt_2}{t_2}\,t^{\Delta_2}_2\int_0^\infty \frac{dt_1}{t_1}\,t^{\Delta_1}_1\,G_{T}(t_1{\hat X}_1,t_2{\hat X}_2) \label{2pt} \\
   &=\left(\frac{m}2\right)^{d-2\Delta_1}\,\frac{C^{\text{flat}}_{\Delta_1}}{(-2Q_1\cdot Q_2+i\epsilon)^{\Delta_1}}(2\pi )\delta(i(\Delta_1-\Delta_2))\,, \nonumber
\end{align}
with normalisation 
\begin{equation}\label{celest2ptnorm}
   C^{\text{flat}}_{\Delta}= \frac{1}{4\pi^{\frac{d+2}2}}\,\Gamma(\Delta)\Gamma(\Delta-\tfrac{d}2).
\end{equation}
Note that the two point function is non-vanishing in the case that one operator is on the past boundary and the other on the future boundary. Such two point functions differ by phase factors which are encoded in the $i\epsilon$ prescription in eq.~\eqref{2pt}:
\begin{multline}
\left\langle\mathcal{O}_{\Delta_1}^{i}(Q_1)\mathcal{O}_{\Delta_2}^{j}(Q_2)\right\rangle\\=\left(\frac{m}2\right)^{d-2\Delta_1}\begin{pmatrix}
1& e^{-i\pi\Delta_1}\\
e^{-i\pi\Delta_1}& 1
\end{pmatrix}\frac{C^{\text{flat}}_{\Delta_1}}{|-2Q_1\cdot Q_2|^{\Delta_1}}\,(2\pi )\delta(i(\Delta_1-\Delta_2))\,.
\end{multline}
This is diagonalised by the following orthogonal operators:
\begin{subequations}
\begin{align}
\mathcal{O}^>_\Delta&=\frac{e^{i\frac{\pi}4(\Delta-1)}}{2}\left(\mathcal{O}_\Delta^--\mathcal{O}_\Delta^+\right),\\
\mathcal{O}^<_\Delta&=\frac{e^{i\frac{\pi}4\Delta}}{2}\left(\mathcal{O}_\Delta^-+\mathcal{O}_\Delta^+\right),
\end{align}
\end{subequations}
with two-point functions:
\begin{subequations}
\begin{align}
\left\langle\mathcal{O}_{\Delta_1}^{>}(Q_1)\mathcal{O}_{\Delta_2}^{>}(Q_2)\right\rangle&=\left(\frac{m}2\right)^{d-2\Delta_1}\sin\left(\tfrac{\pi}2\Delta_1\right)\frac{C^{\text{flat}}_{\Delta_1}}{|-2Q_1\cdot Q_2|^{\Delta_1}}\,(2\pi )\delta(i(\Delta_1-\Delta_2)),\\
\left\langle\mathcal{O}_{\Delta_1}^{<}(Q_1)\mathcal{O}_{\Delta_2}^{<}(Q_2)\right\rangle&=\left(\frac{m}2\right)^{d-2\Delta_1}\cos\left(\tfrac{\pi}2\Delta_1\right)\frac{C^{\text{flat}}_{\Delta_1}}{|-2Q_1\cdot Q_2|^{\Delta_1}}\,(2\pi )\delta(i(\Delta_1-\Delta_2)).
\end{align}
\end{subequations}

\subsection{Contact diagrams}
\label{sec::CCasWD}

In this section we consider consider contact diagrams in a theory of scalar fields $\phi_i$, $i=1,\ldots,n$, which interact through the vertex
\begin{equation}
    {\cal V}\left(X\right) = g \phi_1\left(X\right) \ldots \phi_n\left(X\right).
\end{equation}
At linear order in the coupling $g$, the corresponding $n$-point celestial correlator is given by
\begin{equation}\label{contactelestial}
    \left\langle\mathcal{O}_1(Q_1)\ldots \mathcal{O}_n(Q_n)\right\rangle=-ig\int d^{d+2}X\,G_{\Delta_1}^{\text{flat}}(X,Q_1)\cdots G_{\Delta_n}^{\text{flat}}(X,Q_n)\,.
\end{equation}
Recall that the celestial bulk-to-boundary propagator factorise \eqref{explcelesbubo} into the corresponding (analytically continued) EAdS bulk-to-boundary propagator times the radial kernel of the Kontorovich-Lebedev transform. This makes manifest that celestial contact diagrams are proportional to their corresponding Witten diagrams in EAdS upon integrating out the radial direction. In appendix \ref{appendix::MINKINT}, using the explicit expression \eqref{explcelesbubo} for the Celestial bulk-to-boundary propagators, it is shown how the integral over $X \in \mathbb{M}^{d+2}$ in two different ways: 1. In Cartesian coordinates. 2. In the hyperbolic slicing of $\mathbb{M}^{d+2}$. Both ways give the same factorised result: 
\begin{multline}
    \left\langle\mathcal{O}_1(Q_1)\ldots \mathcal{O}_n(Q_n)\right\rangle=
    \sin \left(\tfrac{\pi}{2}  (\Delta_1+\ldots+\Delta_n-d)\right) \left(\prod_i c^{\text{dS-AdS}}_{\Delta_i}\right) \\ \times R_{\Delta_1\ldots\Delta_n}\left(m_1,\ldots,m_n\right)D_{\Delta_1\ldots\Delta_n}(Q_1,\ldots,Q_n),
\end{multline}
where $D_{\Delta_1\dots\Delta_n}(Q_1,\ldots,Q_n)$ is the familiar EAdS $D$-function \cite{DHoker:1999kzh}, which here is defined over the complexified light-cone $Q_i^2=0$ via the $i \epsilon$ prescription of the celestial bulk-to-boundary propagator (see appendix \ref{appendix::symanzik}). The coefficient $R_{\Delta_1\dots\Delta_n}\left(m_1,\ldots,m_n\right)$ instead arises from the integral over the radial direction and is naturally given by the following Mellin-Barnes integral (see appendix \ref{appendix::MINKINTHYPERBOLIC}):
\begin{equation}\label{radialcontrib}
   R_{\Delta_1\dots\Delta_n}\left(m_1,\ldots,m_n\right)= \int[ds_i]_n (2\pi i)\delta\left(-\frac{d+2}2+\sum_j\left(s_j+\tfrac{d}{4}\right)\right)\prod_i \widetilde{{\cal K}}^{\left(m_i\right)}_{i\left(\frac{d}{2}-\Delta_i\right)}(s_i)\,,
\end{equation}
where $\widetilde{{\cal K}}^{\left(m_i\right)}_{i\left(\frac{d}{2}-\Delta_i\right)}(s_i)$ is the Mellin transform of kernel $\mathcal{K}^{(m)}_{i\left(\frac{d}{2}-\Delta_i\right)}$ of the Kontorovich-Lebedev transform (see Appendix \ref{appendix::KL-transform}). It is interesting to note the similarity between the \eqref{radialcontrib} for the radial integral and the corresponding contact Witten diagram in momentum space (\emph{cf.} equation (3.19) of \cite{Sleight:2021plv}).

\vskip 4pt
Let us make a few comments:

\begin{itemize}
    \item In a previous work \cite{Iacobacci:2022yjo} we instead employed the regularised definition \eqref{regdefcpw} of conformal primary wave functions to show that the corresponding \emph{celestial amplitudes} can be perturbatively re-cast as EAdS Witten diagrams. In the above we have shown that such a relationship with EAdS Witten diagrams also holds for \emph{celestial correlators} \eqref{celestialcorrelationnewdef0} as well, though there are some key differences. In particular, when using the regularised definition \eqref{regdefcpw} of conformal primary wave functions, the contribution \eqref{radialcontrib} from the radial direction contains folded singularities in the mass $m_i$\footnote{By folded singularity we mean singularities when a mass $m_i$ is equal to a linear combination of the other masses. See e.g. equation (4.40a) of \cite{Iacobacci:2022yjo}.} owing to the fact that, according to the definition \eqref{regdefcpw}, incoming and outgoing conformal primary wave functions have different $i \epsilon$ prescriptions. See equation (4.24) of \cite{Iacobacci:2022yjo}. Using the modified prescription \eqref{celestialcorrelationnewdef0}, in which all propagators have the same $i\epsilon$ prescription, the \emph{entire} radial contribution \eqref{radialcontrib} now resembles a momentum space contact Witten diagram in EAdS where the masses $m_i$ play the role of the modulus of the boundary momentum and the Kontorovich-Lebedev kernels are proportional to the momentum space EAdS bulk-to-boundary propagators.\footnote{The Kontorovich-Lebedev kernel \eqref{KL-t} and EAdS bulk-to-boundary propagators in momentum space are given by the same type of Bessel-$K$ function \cite{Gubser:1998bc}. Consistent momentum space EAdS Witten diagrams do not contain folded singularities \cite{Bzowski:2013sza}.} It would be interesting to understand whether there is any significance behind these similarities of the radial contribution with momentum space EAdS Witten diagrams.
    
    \item The overall sinusoidal factor \begin{equation}\label{sinefactor}
     \sin \left(\tfrac{\pi}{2}  (\Delta_1+\ldots+\Delta_n-d)\right),
\end{equation}
arises from combining the contributions (appendix \ref{appendix::MINKINTHYPERBOLIC}) from the regions ${\cal A}_\pm$ and ${\cal D}_\pm$ in the hyperbolic slicing of $\mathbb{M}^{d+2}$, which each differ by a phase. This is reminiscent of a similar result for contact diagram contributions to boundary correlation functions in dS space, which are related to their corresponding EAdS contact Witten diagrams by multiplying the latter by a sinusoidal factor depending on $\Delta_i$, $n$ and $d$. In that case the sinusoidal factor arose from combining the contributions to the dS contact diagram from each branch of the in-in contour \cite{Sleight:2019mgd,Sleight:2019hfp,Sleight:2020obc,Sleight:2021plv}, which have an equal and opposite phase. In the present case the regions ${\cal A}_\pm$ and ${\cal D}_\pm$ are the analogues are the $\pm$ branches of the in-in contour. The factor \eqref{sinefactor} suggests that the celestial contact diagrams are vanishing for certain values of $\Delta_i$, $n$ and $d$.

\item We have shown that celestial contact diagrams are proportional to the corresponding contact Witten diagram in EAdS$_{d+1}$ appropriately analytically continued on the complexified null cone. In other words 
\begin{equation}
    \left\langle\mathcal{O}_1(Q_1)\ldots \mathcal{O}_n(Q_n)\right\rangle = c_{\Delta_1 \ldots \Delta_n}D_{\Delta_1\ldots\Delta_n}(Q_1,\ldots,Q_n),
\end{equation}
where $c_{\Delta_1 \ldots \Delta_n}$ can be thought of as the ratio
\begin{equation}
    c_{\Delta_1 \ldots \Delta_n} = \frac{\lambda^{\text{flat}}_{\Delta_1 \ldots \Delta_n}}{\lambda^{\text{AdS}}_{\Delta_1 \ldots \Delta_n}},
\end{equation}
of the coefficient $\lambda^{\text{AdS}}_{\Delta_1 \ldots \Delta_n}$ of the contact Witten diagram and $\lambda^{\text{flat}}_{\Delta_1 \ldots \Delta_n}$ the coefficient of the celestial contact diagram. As noted in \cite{Iacobacci:2022yjo}, from on-shell factorisation one expects this perturbative relationship between celestial correlators and EAdS Witten diagrams to extend beyond contact diagrams to processes involving particle exchanges. This is proven in the follow-up work \cite{Iacobacci:2024nhw}.
\end{itemize}

{\bf 3pt contact diagrams.} A simple and instructive example is the case of 3pt functions, where (see appendix \ref{appendix::symanzik}):
\begin{multline}
    D_{\Delta_1\Delta_2\Delta_3}(Q_1,Q_2,Q_3)\\=\frac{\lambda^{\text{AdS}}_{\Delta_1\Delta_2\Delta_3}}{(-2Q_1\cdot Q_2+i\epsilon)^{\frac{\Delta_1+\Delta_2-\Delta_3}2}(-2Q_2\cdot Q_3+i\epsilon)^{\frac{\Delta_2+\Delta_3-\Delta_1}2}(-2Q_3\cdot Q_1+i\epsilon)^{\frac{\Delta_3+\Delta_1-\Delta_2}2}}\,.
\end{multline}
Note that the $i\epsilon$ prescription allows to define the above correlator on the complexified light-cone, thus allowing to seamlessly analytically continue from $Q_+$ to $Q_-$ and vice-versa. The coefficient is proportional to the standard 3pt Witten-diagram coefficient (see e.g. equation (131) of \cite{Costa:2014kfa}) and is given by:
\begin{align}
\lambda^{\text{AdS}}_{\Delta_1,\Delta_2,\Delta_3}=-g\frac{\Gamma\left(\tfrac{\Delta_1+\Delta_2+\Delta_3-d}2\right)\Gamma\left(\tfrac{\Delta_1+\Delta_2-\Delta_3}2\right)\Gamma\left(\tfrac{\Delta_2+\Delta_3-\Delta_1}2\right)\Gamma\left(\tfrac{\Delta_3+\Delta_1-\Delta_2}2\right)}{16\pi^d\, \Gamma \left(\Delta_1-\tfrac{d}{2}+1\right) \Gamma \left(\Delta_2-\tfrac{d}{2}+1\right) \Gamma \left(\Delta_3-\tfrac{d}{2}+1\right)}.
\end{align}
By simplifying the $i\epsilon$ prescription one obtains:
{\allowdisplaybreaks\begin{subequations}
\begin{align}\nonumber
    D^{\pm\pm\pm}_{\Delta_1\Delta_2\Delta_3}(Q_1,Q_2,Q_3)&=\frac{\lambda^{\text{AdS}}_{\Delta_1,\Delta_2,\Delta_3}}{|-2Q_1\cdot Q_2|^{\frac{\Delta_1+\Delta_2-\Delta_3}2}|-2Q_2\cdot Q_3|^{\frac{\Delta_2+\Delta_3-\Delta_1}2}|-2Q_3\cdot Q_1|^{\frac{\Delta_3+\Delta_1-\Delta_2}2}}\\&\hspace*{0.5cm}+\text{contact}\,,\\
    D^{\pm\pm\mp}_{\Delta_1\Delta_2\Delta_3}(Q_1,Q_2,Q_3)&=\frac{\lambda^{\text{AdS}}_{\Delta_1,\Delta_2,\Delta_3}e^{-i\pi\Delta_3}}{|-2Q_1\cdot Q_2|^{\frac{\Delta_1+\Delta_2-\Delta_3}2}|-2Q_2\cdot Q_3|^{\frac{\Delta_2+\Delta_3-\Delta_1}2}|-2Q_3\cdot Q_1|^{\frac{\Delta_3+\Delta_1-\Delta_2}2}}\nonumber\\&\hspace*{0.5cm}+\text{contact}\,,\\
    D^{\pm\mp\pm}_{\Delta_1\Delta_2\Delta_3}(Q_1,Q_2,Q_3)&=\frac{\lambda^{\text{AdS}}_{\Delta_1,\Delta_2,\Delta_3}e^{-i\pi\Delta_2}}{|-2Q_1\cdot Q_2|^{\frac{\Delta_1+\Delta_2-\Delta_3}2}|-2Q_2\cdot Q_3|^{\frac{\Delta_2+\Delta_3-\Delta_1}2}|-2Q_3\cdot Q_1|^{\frac{\Delta_3+\Delta_1-\Delta_2}2}}\nonumber\\&\hspace*{0.5cm}+\text{contact}\,
    \\
    D^{\mp\pm\pm}_{\Delta_1\Delta_2\Delta_3}(Q_1,Q_2,Q_3)&=\frac{\lambda^{\text{AdS}}_{\Delta_1,\Delta_2,\Delta_3}e^{-i\pi\Delta_1}}{|-2Q_1\cdot Q_2|^{\frac{\Delta_1+\Delta_2-\Delta_3}2}|-2Q_2\cdot Q_3|^{\frac{\Delta_2+\Delta_3-\Delta_1}2}|-2Q_3\cdot Q_1|^{\frac{\Delta_3+\Delta_1-\Delta_2}2}}\nonumber\\&\hspace*{0.5cm}+\text{contact}\,
\end{align}
\end{subequations}}
where $\pm$ refers to $Q_i \in S^\pm_d$ and where the contact terms are proportional to distributional pieces in the $Q_i$ as $\epsilon\to0$. Similar expressions can be recovered for higher-point functions.

\section{Celestial correlators for Minkowski CFTs}
\label{sec::Conf}

In this work we have proposed a new definition of celestial correlation functions as the Mellin transform of correlation functions in Minkowski space extrapolated to the conformal boundary:
\begin{align}\label{defsect5}
    \left\langle\mathcal{O}_{\Delta_1}(Q_1)\ldots \mathcal{O}_{\Delta_n}(Q_n)\right\rangle =\prod_i \lim_{{\hat Y}_i\to Q_i}\,\int^\infty_0 \frac{dt_i}{t_i}\,t_i^{\Delta_i}\left\langle\phi_1(t_1\hat{Y}_1)\ldots \phi_n(t_n\hat{Y}_n)\right\rangle.
\end{align}
This definition allows to define celestial correlation functions for any theory in Minkowski space, even in cases where the S-matrix does not exist.

\vskip 4pt
CFTs in Minkowski space are important examples of theories where the standard notion of an S-matrix is ill-defined. Despite this, Minkowski CFTs are defined non-perturbatively by conformal symmetry, unitarity and a consistent operator product expansion, which are the three main pillars of the conformal bootstrap programme \cite{Simmons-Duffin:2016gjk,Poland:2018epd}. This is to be contrasted with our understanding of the properties of celestial correlation functions and how they encode consistent Minkowski bulk physics. Taking the bulk Minkowski theory to be a CFT, the definition \eqref{defsect5} provides an opportunity to study the properties of celestial correlation functions from their relatively well understood Minkowski CFT counterparts!

\vskip 4pt
Three and four-point conformal correlation functions of quasi-primary fields $\phi_{{\bar \Delta}_i}$ in $\mathbb{M}^{d+2}$ are constrained by conformal symmetry to take the following form \cite{Polyakov:1970xd}
\begin{subequations}\label{MINKCFT3pt4pt}
 \begin{align}
\left\langle \phi_{{\bar \Delta}_1}(X_1)\phi_{{\bar \Delta}_2}(X_2)\phi_{{\bar \Delta}_3}(X_3)\right\rangle&=\frac{f_{{\bar \Delta}_1,{\bar \Delta}_2,{\bar \Delta}_3}}{(X_{12}^2)^{\frac{{\bar \Delta}_1+{\bar \Delta}_2-{\bar \Delta}_3}2}(X_{23}^2)^{\frac{{\bar \Delta}_2+{\bar \Delta}_3-{\bar \Delta}_1}2}(X_{31}^2)^{\frac{{\bar \Delta}_3+{\bar \Delta}_1-{\bar \Delta}_2}2}}\,,\\
\left\langle \phi_{{\bar \Delta}_1}(X_1)\phi_{{\bar \Delta}_2}(X_2)\phi_{{\bar \Delta}_3}(X_3)\phi_{{\bar \Delta}_4}(X_4)\right\rangle&=\frac{1}{(X_{12}^2)^{\frac{{\bar \Delta}_1+{\bar \Delta}_2}2}(X_{34}^2)^{\frac{{\bar \Delta}_3+{\bar \Delta}_4}2}}\left(\frac{X_{24}^2}{X_{14}^2}\right)^{\frac{{\bar \Delta}_1-{\bar \Delta}_2}2}\left(\frac{X_{14}^2}{X_{13}^2}\right)^{\frac{{\bar \Delta}_3-{\bar \Delta}_4}2}\,f(u,v),
\end{align}
\end{subequations}
where $f_{{\bar \Delta}_1,{\bar \Delta}_2,{\bar \Delta}_3}$ is a constant and $f(u,v)$ a function of the usual conformal invariant cross ratios $u$ and $v$. We use the notation $X^2_{ij} = \left(X_i-X_j\right)^2$. To plug these into the definition \eqref{defsect5} of celestial correlation functions we take the light-cone limit $X_i \to t_i Q_i$. In this limit the we have $X^2_{ij} \to -2 t_i t_j Q_i \cdot Q_j$, which are homogeneous in the $t_i$ and simplifies significantly the the Mellin transform \eqref{defsect5}, which reduce to the integrals\footnote{The Dirac delta function function should be considered as a distribution in complex space, as was done in \cite{Sleight:2019hfp} by regularising the integral depending on the location of the integration contour.}
\begin{align}
\int_0^\infty \frac{dt_i}{t_i}\,t_i^{\Delta_i-{\bar \Delta}_i}=2\pi\delta(i(\Delta_i-{\bar \Delta}_i))\,.
\end{align}
The celestial correlation functions corresponding to the Minkowski conformal three- and four-point functions \eqref{MINKCFT3pt4pt} are therefore
\begin{multline}
\left\langle O_{\Delta_1}(Q_1)O_{\Delta_2}(Q_2)O_{\Delta_3}(Q_3)\right\rangle \\=  \frac{f_{\Delta_1,\Delta_2,\Delta_3}\prod\limits^3_{i=1} 2\pi \delta(i(\Delta_i-{\bar \Delta}_i))}{(X_{12}^2)^{\frac{\Delta_1+\Delta_2-\Delta_3}2}(X_{23}^2)^{\frac{\Delta_2+\Delta_3-\Delta_1}2}(X_{31}^2)^{\frac{\Delta_3+\Delta_1-\Delta_2}2}}\Bigg|_{X_{ij}^2\to-2Q_i\cdot Q_j+i\epsilon}\,,
\end{multline}
\begin{align}
\left\langle O_{\Delta_1}(Q_1)O_{\Delta_2}(Q_2)O_{\Delta_3}(Q_3)O_{\Delta_4}(Q_4)\right\rangle&=\frac{\prod\limits^4_{i=1} 2\pi \delta(i(\Delta_i-{\bar \Delta}_i))}{(X_{12}^2)^{\frac{\Delta_1+\Delta_2}2}(X_{34}^2)^{\frac{\Delta_3+\Delta_4}2}}\\&\times\left(\frac{X_{24}^2}{X_{14}^2}\right)^{\frac{\Delta_1-\Delta_2}2}\left(\frac{X_{14}^2}{X_{13}^2}\right)^{\frac{\Delta_3-\Delta_4}2}\,f(u,v)\Bigg|_{X_{ij}^2\to-2Q_i\cdot Q_j+i\epsilon}\,,\nonumber
\end{align}
where $u$ and $v$ are now the cross-ratios in $d$-dimensions. These are Euclidean conformal correlators in $d$ dimensions and the $i\epsilon$ prescription inherited from the Bulk Minkowski CFT allows to distinguish in and out operators depending on whether $Q_i\in S_d^{\pm}$. 

\vskip 4pt
Let us consider a simple concrete example. The mean field theory correlation function of an operator $\phi$ with scaling dimension $\Delta$ is given by
\begin{align}
\left\langle \phi(Q_1)\phi(Q_2)\phi(Q_3)\phi(Q_4)\right\rangle=\frac1{(X_{12}^2X_{34}^2)^{\Delta}}\left(1+u^{\Delta}+\left(\frac{u}{v}\right)^{\Delta}\right)\,.
\end{align}
The corresponding celestial correlation function is
\begin{multline}
\left\langle O_{\Delta_1}^+(Q_1)O_{\Delta_2}^+(Q_2)O_{\Delta_3}^-(Q_3)O_{\Delta_4}^-(Q_4)\right\rangle\\=\frac{\prod\limits^4_{i=1} 2\pi\delta(i(\Delta_i-\Delta))}{(y_{12}^2y_{34}^2)^{\Delta}}\left(1+e^{-2\pi i\Delta}u^\Delta+e^{-2\pi i\Delta}\left(\frac{u}{v}\right)^{\Delta}\right)\,,
\end{multline}
where the $y_i$ are $d$-dimensional Euclidean vectors and we took $Q_{1,2}\in S^+_d$ and $Q_{3,4}\in S^-_d$. The phase factors $e^{-2\pi i\Delta}$ originate from the bulk $i \epsilon$ prescription and thus encode bulk Minkowski causality. These examples could therefore be useful toy models to gain a better understanding of celestial CFTs and how they encode bulk unitarity and causality. We leave a detailed analysis of the properties of the above celestial correlators for the future!

\section{Conclusions}\label{sec::Intro}

We conclude with some future directions: 

\begin{itemize}
\item While our prescription for celestial correlation functions holds both for massive and massless particles, it would be desirable to study in the massless case in more detail. In this case, the conformal boundary can be reached on-shell, thus suggesting that there should be a relation between our prescription and the $S$-matrix for massless particles. It would be very interesting to see if the two prescriptions for celestial correlators can be related and how the S-matrix is embedded within celestial correlators \eqref{celestialcorrelationnewdef0}.

\item Independently of the above point, we note that it is possible to put external legs on-shell by considering the discontinuity of the Feynman propagator:
\begin{align}
    \text{Disc}_{P^2}\left[\frac1{P^2+m^2-i\epsilon}\right]=\Im\left[\frac1{P^2+m^2-i\epsilon}\right]=(2\pi i)\delta(P^2+m^2)\,.
\end{align}
Taking the Mellin transform of the above equation one thus end up with a consistent regularisation for the Mellin transform of the on-shell condition:
\begin{align}
\text{disc}\left[G_\Delta^{\text{flat}}(X,Q)\right]=\Im\left[\lim_{\hat{Y}\to Q}\int_0^\infty\frac{dt}{t}\,t^\Delta\,G_T(X,t\hat{Y})\right]\,.
\end{align}
At the level of celestial correlation functions this is equivalent to taking the discontinuity with respect to all external legs:
\begin{multline}
\left\langle\mathcal{O}_{\Delta_1}(Q_1)\ldots \mathcal{O}_{\Delta_n}(Q_n)\right\rangle\Big|_{\text{on-shell}}\\=\prod_i\Im_i\left[\prod_i \lim_{{\hat Y}_i\to Q_i}\,\int^\infty_0 \frac{dt_i}{t_i}\,t_i^{\Delta_i}\left\langle\phi_1(t_1\hat{Y}_1)\ldots \phi_n(t_n\hat{Y}_n)\right\rangle\right],
\end{multline}
which is related to the S-matrix via the LSZ formula. This might elucidate the relation between the prescription proposed in this work and the S-matrix.

\item An advantage of this modified definition of celestial correlators is that it applies to theories which do not have an S-matrix. An important example of such theories are Minkowski CFTs. A more detailed study of our prescription in the case that the bulk theory is a Minkowski CFT would be relevant to understand non-perturbative properties of the exotic Euclidean CFTs arising at the boundary of Minkowski space, such as unitarity, causality and analyticity and how they are encoded. From the simple examples we have considered it is already apparent that the Celestial correlators dual to Minkowski CFTs are distributions and it would be interesting to clarify their analyticity properties more generally.

\item Note that, taking $Q_{1,2} \in S^{+}_d$ and $Q_{3,4} \in S^{-}_d$ the definition \eqref{celestialcorrelationnewdef0} of celestial four-point functions is such that the corresponding singularities corresponds to the Regge limit of the bulk Minkowski correlator. It would be interesting to clarify the interplay between bulk Minkowski correlator, their Regge limit and Celestial correlators.

\item The fact that celestial correlation functions can be perturbatively recast as Witten diagrams in EAdS implies that they have the same analytic structure and in particular admit a conformal partial wave decomposition. Assuming that this continues to hold at the non-perturbative level, it was noted in \cite{Iacobacci:2022yjo} that unitarity implies a non-perturbative positivity constraint on the spectral density of celestial four-point functions - generalising the same observation \cite{Hogervorst:2021uvp,DiPietro:2021sjt} for dS boundary correlators to any unitary Euclidean CFT. It would be interesting use this to derive non-perturbative constraints on bulk Minkowski physics.

\item Our prescription naturally extends to the case of spinning fields. While the extension to massive spinning fields looks straightforward, in the massless case it will be important to study BMS symmetries for the celestial correlators so defined, derive the corresponding Ward identities and identify soft degrees of freedom at the boundary e.g. along the lines of \cite{Donnay:2018neh,Donnay:2020guq,Donnay:2022sdg}. 
\end{itemize}

\section*{Acknowledgments}

The research of CS was partially supported by the STFC grant ST/T000708/1. The research of MT was partially supported by the INFN initiative STEFI.

\newpage

\begin{appendix}

\section{The Kontorovich-Lebedev transform}\label{appendix::KL-transform}

A complete orthogonal basis for elements of $L^2\left(\mathbb{R}^+,dR R^{d-1}\right)$ are given by\footnote{Note that here we use a different normalisation to our previous work \cite{Iacobacci:2022yjo} including a factor of $m$ which allows to define a massless limit while preserving the orthogonality and completeness.} 
\begin{align}\label{KL-t}
    \mathcal{K}^{(m)}_{\alpha}(R)=\left\langle R|\tilde{K}^{(m)}_\alpha\right\rangle=\left(\frac{m}2\right)^{-i\alpha}\frac{2R^{-d/2}}{\Gamma(-i\alpha)}\,K_{i\alpha}(m R)\,,
\end{align}
in terms of the Bessel-K function. See \cite{Iacobacci:2022yjo} for a derivation of completeness and orthogonality relations. The decomposition of an element of $L^2\left(\mathbb{R}^+,dR R^{d-1}\right)$ is implemented by the Kontorovich-Lebedev transform 
\begin{subequations}
 \begin{align}
    \phi(R,{\hat X}) &= \frac{1}{2}\int^{\frac{d}{2}+i\infty}_{\frac{d}{2}-i\infty}\frac{d\Delta}{2\pi i}\phi_{\Delta}({\hat X}){\cal K}^{(m)}_{i\left(\frac{d}{2}-\Delta\right)}\left(R\right),\\
  \phi_{\Delta}({\hat X}) &= \int^\infty_0 dR\, \phi(R,{\hat X})\left[ {\cal K}^{(m)}_{i\left(\frac{d}{2}-\Delta\right)}\left(R\right)\right]^*.
\end{align}   
\end{subequations}
This provides a map between fields $\phi\left(X\right)$ living on $\mathbb{M}^{d+2}$ and fields $\phi_{\Delta}({\hat X})$ living on the $H^+_{d+1}$ hypersurface, and vice versa.

\vskip 4pt
The kernel $\mathcal{K}^{(m)}_{\alpha}(R)$ of the Kontorovich-Lebedev transform is most conveniently expressed in terms of its Mellin transform $\widetilde{ \mathcal{K}}_\alpha^{(m)}(s)$:
\begin{subequations}
 \label{KL-mellin}
\begin{align}
\mathcal{K}_\alpha^{(m)}(R)&=\int_{-i\infty}^{+i\infty}\frac{ds}{2\pi i}\,\,\widetilde{ \mathcal{K}}_\alpha^{(m)}(s)R^{-d/2-2s},\\
\widetilde{ \mathcal{K}}_\alpha^{(m)}(s)&=\left(\frac{m}2\right)^{-i\alpha}\frac{1}{\Gamma(-i\alpha)}\,\Gamma\left(s+\tfrac{i\alpha}2\right)\Gamma\left(s-\tfrac{i\alpha}2\right)\left(\frac{m}2\right)^{-2s}.
\end{align}  
\end{subequations}

\section{Conformal Primary Wave functions}
\label{A::CPWFs}

In this appendix we derive the Fourier transform of conformal primary wave functions
\begin{align}
    \tilde{\phi}^\pm_{\Delta}(P;Q)=\int d^{d+2}X\,\phi^\pm_\Delta(X,Q)e^{-iP\cdot X}\,,
\end{align}
 which were defined in \cite{Pasterski:2016qvg,Pasterski:2017kqt} via the regularisation
\begin{align}\label{regdefcpwA}
  \hspace*{-0.25cm}  \phi^\pm(X,Q)=\frac{1}{2\pi^{\frac{d+2}{2}}}\left(\frac{m}{2}\right)^{\frac{d}{2}-\Delta}\frac{\Gamma(\Delta)}{(-2X\cdot Q\mp i\epsilon)^\Delta} (\sqrt{X^2\mp i\epsilon})^{\Delta-\frac{d}{2}} {K}_{\Delta-\tfrac{d}2}(m\sqrt{X^2\mp i\epsilon})\,.
\end{align}
To this end, we reduce the Minkowski integral to a Gaussian integral by introducing two Schwinger parameters:
\begin{align} \nonumber
    \tilde{\phi}^\pm_{\Delta}(P;Q)&= \frac{1}{4 \pi^{\frac{d+2}{2}}}\left(\frac{m}{2}\right)^{\frac{d}{2}-\Delta}\frac1{\Gamma(\Delta)}\int^{+i\infty}_{-i\infty}\frac{ds}{2\pi i}\Gamma(s+\tfrac12(\Delta-\tfrac{d}{2}))\left(\frac{m}2\right)^{-2s}(\pm i)^{-\Delta}\nonumber\\ \nonumber &\times \int^{+i\infty}_{-i\infty}\frac{ds}{2\pi} \Gamma\left(s+\tfrac{1}{2}\left(\Delta-\tfrac{d}{2}\right)\right)\left(\mp i\right)^{-s+\tfrac{1}{2}\left(\Delta-\frac{d}{2}\right)}\\ \nonumber
    & \times \int_0^\infty\frac{dt}{t}\,t^\Delta  \int_0^\infty\frac{du}{u}\,u^{s-\frac{1}{2}\left(\Delta-\frac{d}{2}\right)} \\
    & \times \int d^{d+2}X\,e^{\pm i 2 t X \cdot Q \mp i \frac{u m^2 X^2}{4} - i P \cdot X}.
\end{align}
Evaluating the Gaussian integral gives
\begin{align} \nonumber
    \tilde{\phi}^\pm_{\Delta}(P;Q)&= \frac{1}{4} \left(\frac{m}{2}\right)^{\frac{d}{2}-\Delta} \left(\pm i\right)^{\Delta-\frac{d}{2}} \int^{+i\infty}_{-i\infty}\frac{ds}{2\pi} \Gamma\left(s+\tfrac{1}{2}\left(\Delta-\tfrac{d}{2}\right)\right)\left(\mp i\right)^{-s+\tfrac{1}{2}\left(\Delta-\frac{d}{2}\right)}\left(\frac{m^2}{4}\right)^{-s+\tfrac{1}{2}\left(\Delta-\frac{d}{2}\right)}\\
    & \times \int_0^\infty\frac{dt}{t}\,t^\Delta  \int_0^\infty\frac{du}{u}\,u^{-s+\frac{1}{2}\left(\Delta-\frac{d}{2}\right)+\frac{d+2}{2}}  e^{\pm \frac{iu}{4}\left(P \pm 2t Q\right)^2}.
\end{align}
Performing the $t$-integral one gets:
\begin{align} \nonumber
    \tilde{\phi}^\pm_{\Delta}(P;Q)&= \frac{1}{4} \left(\frac{m}{2}\right)^{\frac{d}{2}-\Delta} \left(\pm i\right)^{\Delta-\frac{d}{2}} \frac{\Gamma\left(\Delta\right)}{\left(i P \cdot Q + \epsilon\right)^{\Delta}} \\ \nonumber 
    & \times \int^{+i\infty}_{-i\infty}\frac{ds}{2\pi} \Gamma\left(s+\tfrac{1}{2}\left(\Delta-\tfrac{d}{2}\right)\right)\left(\mp i\right)^{-s+\tfrac{1}{2}\left(\Delta-\frac{d}{2}\right)}\left(\frac{m^2}{4}\right)^{-s+\tfrac{1}{2}\left(\Delta-\frac{d}{2}\right)}\\
    & \times  \int_0^\infty\frac{du}{u}\,u^{-s-\frac{1}{2}\left(\Delta-\frac{d}{2}\right)+1}  e^{\pm \frac{iu}{4}P^ 2},
\end{align}
And finally for the $u$ integral:
\begin{align} \nonumber
    \tilde{\phi}^\pm_{\Delta}(P;Q)&= \pm i\frac{\Gamma\left(\Delta\right)}{\left(i P \cdot Q + \epsilon\right)^{\Delta}}\underbrace{\frac{1}{m^2} \int^{+i\infty}_{-i\infty}\frac{ds}{2\pi} \Gamma\left(s\right)\Gamma\left(1-s\right)\left(\frac{m^2}{P^2\pm i\epsilon}\right)^{1-s}}_{\frac{1}{P^2+m^2\pm i \epsilon}}.
\end{align}
This expression is to be compared with the extrapolation of the (anti-)time-ordered propagator, 
\begin{align}
    G_T(X,Y)&=\int \frac{d^{d+2}P}{(2\pi)^{d+2}}\frac{-i}{P^2+m^2-i\epsilon}\,e^{iP\cdot(X-Y)},\\
    G_{\bar T}(X,Y)&=\int \frac{d^{d+2}P}{(2\pi)^{d+2}}\frac{+i}{P^2+m^2+i\epsilon}\,e^{iP\cdot(X-Y)}.
\end{align}
Focusing on the Feynman propagator $G_T(X,Y)$, we have
\begin{align}
    \lim_{{\hat Y}\to Q}\int^\infty_0 \frac{dt}{t} t^\Delta \, G_T(X,t{\hat Y}) & = \int \frac{d^{d+2}P}{(2\pi)^{d+2}}\frac{-i}{P^2+m^2-i\epsilon}\,e^{iP\cdot X} \int^\infty_0 \frac{dt}{t} t^\Delta e^{-i Y \cdot t Q}\\
    & = \int \frac{d^{d+2}P}{(2\pi)^{d+2}}\frac{-i}{P^2+m^2-i\epsilon}\frac{\Gamma\left(\Delta\right)}{\left(i P \cdot Q + \epsilon\right)^{\Delta}}\,e^{iP\cdot X},
\end{align}
which corresponds to $\tilde{\phi}^-_{\Delta}(P;Q)$ above. For the anti-time-ordered $G_{\bar T}(X,Y)$ we similarly have 
\begin{align}
    \lim_{{\hat Y}\to Q}\int^\infty_0 \frac{dt}{t} t^\Delta \, G_{\hat T}(X,t{\hat Y})  = \int \frac{d^{d+2}P}{(2\pi)^{d+2}}\frac{+i}{P^2+m^2-i\epsilon}\frac{\Gamma\left(\Delta\right)}{\left(i P \cdot Q + \epsilon\right)^{\Delta}}\,e^{iP\cdot X},
\end{align}
which corresponds to $\tilde{\phi}^+_{\Delta}(P;Q)$.

\vskip 4pt
This shows that the conformal primary wave functions defined via the regularisation \eqref{regdefcpwA} correspond to off-shell (anti)-time-ordered propagators.

\section{Celestial bulk-to-boundary propagators}
\label{appendix::CBUBO}

In this appendix we derive the closed form expression \eqref{explcelesbubo} for the celestial bulk-to-boundary propagators. We start from the definition
\begin{equation}\label{celestialbuboA}
    G^{\text{flat}}_{\Delta}\left(X,Q\right) = \int^\infty_0 \frac{dt}{t} t^\Delta\,G_T\left(X,tQ\right),
\end{equation}
which is simply \eqref{celestialbubo} but commuting the limit with the integral in $t$, where the $i\epsilon$ prescription in the Feynman propagator which ensures that the integral converges exponentially. In the ``mostly plus" signature the Feynman propagator is given by:
\begin{align}
    G_T(X,Y)=\int \frac{d^{d+2}P}{(2\pi)^{d+2}}\frac{-i}{P^2+m^2-i\epsilon}\,e^{-iP\cdot(X-Y)}.
\end{align}
It is convenient to first evaluate the momentum integral, which can be done employing Schwinger parameterisation:
\begin{equation}\label{SPFeyn}
    G_{T}(X,Y) = \int^\infty_0 \frac{du}{u} u \int \frac{d^{d+2}P}{\left(2\pi\right)^{d+2}} \exp\left[-iu\left(P^2+m^2\right)-i P \cdot \left(X-Y\right)\right].
\end{equation}
This reduces the momentum integral to a Gaussian integral, giving
\begin{equation}
    G_{T}(X,Y) = \frac{1}{2^{d+2}} \frac{1}{\pi^{\frac{d+2}{2}}} i^{-\frac{d}{2}}\int^\infty_0 \frac{du}{u} u^{-\frac{d}{2}}   \exp\left[-\frac{i}{4u}\left(X-Y\right)^2-i u m^2\right].
\end{equation}
Plugging into the extrapolation formula \eqref{celestialbuboA} for the bulk-to-boundary propagator we have
\begin{align}
    G^{\text{flat}}_{\Delta}\left(X,Q\right) & = i^{-\frac{d}{2}} \frac{1}{2^{d+2}} \frac{1}{\pi^{\frac{d+2}{2}}} \int^\infty_0 \frac{dt}{t} t^{\Delta} \exp\left[-\frac{i}{2} t X \cdot Q\right]\\ & \hspace*{3cm}\times \int^\infty_0 \frac{du}{u} u^{\Delta-\frac{d}{2}} \exp\left[\frac{i}{4u} X^2 - iu m^2\right]. \nonumber 
\end{align}
One then obtains the expression
\begin{align} 
    G^{\text{flat}}_{\Delta}(X;Q)&=\frac{\Gamma(\Delta)}{4\pi^{\frac{d+2}2}}\left(\frac{m}2\right)^{\frac{d}{2}-\Delta}\,\frac{(X^2+i\epsilon)^{\frac12(\Delta-\frac{d}{2})}}{\left(-2 X\cdot Q+i\epsilon\right)^\Delta}\\\nonumber&\qquad\times\int_{-i\infty}^{+i\infty}\frac{ds}{2\pi i}\,\Gamma (s+\tfrac12(\Delta-\tfrac{d}{2}))\Gamma(s-\tfrac12(\Delta-\tfrac{d}{2}))
    \left(\tfrac{m}2\sqrt{{X^2}+i\epsilon}\right)^{-2s}\\
   &=\frac{1}{2\pi^{\frac{d+2}2}}
    \,\frac{\Gamma(\Delta)}{(-2 X\cdot Q+i\epsilon)^\Delta}\,\left(\sqrt{X^2+i\epsilon}\right)^{\Delta-\frac{d}{2}}\left(\frac{m}{2}\right)^{\frac{d}{2}-\Delta}K_{\Delta-\frac{d}{2}}(m\sqrt{X^2+i\epsilon})\,, \nonumber
\end{align}
by using the Cahen-Mellin integral
\begin{equation}
    e^{-ium^2} = \int^{+i\infty}_{-i\infty} \frac{ds}{2\pi} \Gamma\left(s\right) \left(ium^2\right)^{-s},   
\end{equation}
and
\begin{align}
    \int^\infty_0 \frac{dt}{t} t^{\Delta} \exp\left[-\frac{i}{2} t X \cdot Q\right] & =  \frac{\Gamma\left(\Delta\right)}{\left(\frac{i}{2} X \cdot Q + i\epsilon\right)^{\Delta}},\\
    \int^\infty_0 \frac{du}{u} u^{\frac{d}{2}-\Delta+s} \exp\left[-\frac{i}{2} u X^2\right] & = \Gamma\left(\frac{d}{2}-\Delta+s\right)\left(-\frac{i}{4}X^2+\epsilon\right)^{\Delta-\frac{d}{2}-s}.
\end{align}
The expression \eqref{explcelesbubo} for the celestial bulk-to-boundary propagator is then obtained by comparing the Bessel-$K$ function $K_{\Delta-\frac{d}{2}}\left(z\right)$ with the kernel \eqref{KL-t} of the Kontorovich-Lebedev transform.

\section{Celestial two-point function}
\label{sec::celestial2pt}

In this appendix we give the derivation for the normalisation \eqref{celest2ptnorm} of the two-point function on the celestial sphere. This is obtained from the Feynman propagator by extrapolating both bulk points to the celestial sphere according to the prescription \eqref{celestialcorrelationnewdef0}:
\begin{align}
 \langle \mathcal{O}_{\Delta_1}\left(Q_1\right) \mathcal{O}_{\Delta_2}\left(Q_2\right) \rangle &= \lim_{{\hat X}_i \to Q_i}\int_0^\infty \frac{dt_2}{t_2}\,t^{\Delta_2}_2\int_0^\infty \frac{dt_1}{t_1}\,t^{\Delta_1}_1\,G_{T}(t_1{\hat X}_1,t_2{\hat X}_2).
 \end{align}
Employing the Schwinger parameterisation \eqref{SPFeyn} we have:
\begin{align}
    \langle \mathcal{O}_{\Delta_1}\left(Q_1\right) \mathcal{O}_{\Delta_2}\left(Q_2\right) \rangle
    &= \frac{i^{-\frac{d}{2}}}{2^{d+2} \pi^{\frac{d+2}{2}}} \int_0^\infty \frac{dt_2}{t_2}\,t^{\Delta_2-\Delta_1}_2\int_0^\infty \frac{dt_1}{t_1}\,t^{\Delta_1}_1\,\exp\left[-\frac{i}{2} t_1 Q_1 \cdot Q_2\right] \\ & \hspace{4cm}  \times \int^\infty_0 \frac{du}{u} u^{-\frac{d}{2}+\Delta_1} \exp\left[-i u m^2\right]. \nonumber
\end{align}
The expression \eqref{2pt}
\begin{multline}
    \langle \mathcal{O}_{\Delta_1}\left(Q_1\right) \mathcal{O}_{\Delta_2}\left(Q_2\right) \rangle = \frac{1}{4 \pi^{\frac{d}{2}+1}} \Gamma\left(\Delta_1\right)\Gamma\left(\Delta_1-\frac{d}{2}\right) \left(\frac{m}{2}\right)^{d-2\Delta_1} \\ \times 2\pi i\, \delta\left(\Delta_1-\Delta_2\right) \frac{1}{\left(-2 Q_1 \cdot Q_2 + i \epsilon \right)^{\Delta_1}},
\end{multline}
is then obtained using that
\begin{align}
\int_0^\infty   \frac{dt_2}{t_2}\,t^{\Delta_2-\Delta_1}_2 &= 2\pi \delta\left(i\left(\Delta_1-\Delta_2\right)\right),\\
    \int^\infty_0 \frac{du}{u} u^{-\frac{d}{2}+\Delta_1} \exp\left[-i u m^2\right] & = i^{\frac{d}{2}-\Delta_1}\Gamma\left(\Delta_1-\frac{d}{2}\right) m^{d-2\Delta_1},\\
    \int_0^\infty \frac{dt_1}{t_1}\,t^{\Delta_1}_1\,\exp\left[-\frac{i}{2} t_1 Q_1 \cdot Q_2\right] &= i^{\Delta_i}\frac{\Gamma\left(\Delta_1\right)}{\left(-\frac{1}{2} Q_1 \cdot Q_2 + i\epsilon \right)^{\Delta_1}}.
\end{align}

Note that the case in which the two operators are shadow requires more careful treatment to capture distributional terms.

\section{Symanzik star formula}
\label{appendix::symanzik}

In this appendix we review the Symanzik star formula for the $D$-function
\begin{equation}\label{DfunctA}
    D_{\Delta_1 \ldots \Delta_n}\left(Q_1, \ldots, Q_n\right) = -g \int_{\text{EAdS}}d^{d+1}{\hat X}\,G^{\text{AdS}}_{\Delta_1}\left({\hat X},Q_1\right) \ldots G^{\text{AdS}}_{\Delta_n}\left({\hat X},Q_n\right),
\end{equation}
which is defined as the $n$-point non-derivative contact Witten diagram in EAdS$_{d+1}$ (\cite{DHoker:1999kzh} appendix A). By now it is well known that upon evaluating the bulk integral the $D$-function can be expressed in the form (see e.g. \cite{Penedones:2010ue,Paulos:2011ie}):
\begin{equation}
    D_{\Delta_1 \ldots \Delta_n}\left(Q_1, \ldots, Q_n\right) = -g\pi^{d/2} \Gamma\left(\frac{-d+\sum_i \Delta_i}{2}\right) \prod_{i} \frac{C^{\text{AdS}}_{\Delta_i}}{\Gamma\left(\Delta_i\right)}\, \int_0^\infty\frac{dt_i}{t_i}t_i^{\Delta_i}\,e^{P^2},
\end{equation}
where 
\begin{align}
    P=\sum_i t_i Q_i&\,,& Q_i^2&=0\,,& Q^0_i&>0\,.
\end{align}
The remaining integrals over the Schwinger parameters $t_i$ can be recast as the Symanzik star formula (see \cite{Paulos:2011ie} appendix B):
\begin{align}
   \prod_{i} \int_0^\infty\frac{dt_i}{t_i}t_i^{\Delta_i}\,e^{P^2}=\left(\prod_{i<j}\int_{-i\infty}^{+i\infty}\frac{ds_{ij}}{2\pi i}\,\Gamma(s_{ij}) \right)\left[\prod_{j}(2\pi i)\delta(\sum_{i\neq j} s_{ij}-\Delta_j)\right](-2Q_i\cdot Q_j)^{-s_{ij}},
\end{align}
giving an integral representation of the D-function \eqref{DfunctA}. The formula follows as a consequence of the Mellin-Barnes representation of the exponential function together the identity:
\begin{align}
    \int_0^\infty \frac{dt}{t}\,t^\alpha=2\pi\delta(i\alpha).
\end{align}

The Symanzik star formula can be analytically continued to the complexified null-cone $Q_i^2=0$ according to the $i\epsilon$ prescription of the celestial bulk-to-boundary propagator \eqref{explcelesbubo}, which originates from that of the Feynman propagator. This gives the more general identity: 
\begin{multline}
   e^{-i\frac{\pi}{4}(\Delta_1+\ldots+\Delta_n)}\prod_{i} \int_0^\infty\frac{dt_i}{t_i}t_i^{\Delta_i}\,e^{-iP^2}\\=\left(\prod_{i<j}\int_{-i\infty}^{+i\infty}\frac{ds_{ij}}{2\pi i}\,\Gamma(s_{ij}) \right)\left[\prod_{j}(2\pi i)\delta(\sum_{i\neq j} s_{ij}-\Delta_j)\right](-2Q_i\cdot Q_j+i\epsilon)^{-s_{ij}}\,,
\end{multline}
which allows to define the analytic continuation of the $D$-function on the complexified null cone induced by celestial contact diagrams \eqref{contactelestial}:
\begin{multline}\label{Danalcont}
    D_{\Delta_1 \ldots \Delta_n}\left(Q_1, \ldots, Q_n\right) = -g \pi^{d/2} \Gamma\left(\frac{-d+\sum_i \Delta_i}{2}\right) \prod_{i} \frac{C^{\text{AdS}}_{\Delta_i}}{\Gamma\left(\Delta_i\right)} \\ \times \left(\prod_{i<j}\int_{-i\infty}^{+i\infty}\frac{ds_{ij}}{2\pi i}\,\Gamma(s_{ij}) \right)\left[\prod_{j}(2\pi i)\delta(\sum_{i\neq j} s_{ij}-\Delta_j)\right](-2Q_i\cdot Q_j+i\epsilon)^{-s_{ij}}.
\end{multline}
Note that the product of delta functions can be recast in a matrix form:
\begin{align}
    \prod_{j}(2\pi i)\delta(\sum_{i\neq j} s_{ij}-\Delta_j)=(2\pi i)^n\delta^{(n)}(M \vec{t}-\vec{\alpha})=(2\pi i)^n\frac{\delta^{(n)}(\vec{t}-M^{-1}\vec{\alpha})}{\det M}\,,
\end{align}
where $\vec{t}$ are the set of $n$ Mellin invariants we will solve for and $\vec{s}$ the set of $\frac{n(n-3)}2$ independent leftover variables. One can then write:
\begin{align}
   \prod_{i} \int_0^\infty\frac{dt_i}{t_i}t_i^{\Delta_i}\,e^{-iP^2}=\frac1{(2\pi i)^{n(n-3)/2}\det M}\int_{-i\infty}^{+i\infty}d\vec{s}\,\Gamma(s_{ij}\left(\vec{s}\right))(2iQ_i\cdot Q_j+\epsilon)^{-s_{ij}\left(\vec{s}\right)}\,.
\end{align}

\section{Minkowski integral for Celestial contact diagrams}
\label{appendix::MINKINT}

In this appendix we evaluate the integral over Minkowski space appearing in contact diagram contributions to celestial correlators \eqref{contactelestial}, which takes the form:
\begin{align}
    I_{\mathbb{M}_{d+2}}&=\int_{\mathbb{M}^{d+2}}\,d^{d+2}X\prod_{i=1}^n G(\hat{X}_\epsilon,Q_i)\mathcal{K}^{(m_i)}_{i\left(\frac{d}{2}-\Delta_i\right)}(\sqrt{X^2+i\epsilon})\\&=\mathcal{N}\int_{\mathbb{M}^{d+2}}\,d^{d+2}X\prod_{i=1}^n\frac{\left(\sqrt{X^2+i\epsilon}\right)^{\Delta_i-\frac{d}{2}}}{(2iX\cdot Q_i+\epsilon)^{\Delta_i}}\,2K_{\left(\Delta_i-\frac{d}{2}\right)}\left(m\sqrt{X^2+i\epsilon}\right),
\end{align}
where the constant ${\cal N}$ is defined by the second equality:
\begin{align}\label{constN}
    \mathcal{N}=\prod_{i}\frac{c^{\text{dS-AdS}}_{\Delta_i}C^{\text{AdS}}_{\Delta_i}}{i^{\Delta_i}\Gamma(\frac{d}{2}-\Delta_i)}\left(\frac{m_i}2\right)^{\frac{d}{2}-\Delta_i}.
\end{align}
In the following sections we perform this integral in two ways: 1. by directly evaluating the Minkowski integral in Cartesian coordinates 2. In the hyperbolic slicing of Minkowski space.

\subsection{In Cartesian coordinates}
The Minkowski integral can be evaluated directly by reducing it to a Gaussian integral. This is easily done employing a Schwinger parameterisation combined with the Mellin-Barnes representation \eqref{KL-mellin} of the kernel $\mathcal{K}^{(m)}_{\Delta-\frac{d}{2}}$ of the Kontorovich-Lebedev transform:
\begin{align}\label{IMINK}
I_{\mathbb{M}_{d+2}}&=\mathcal{N}\int[ds_i]_n \prod_i\,\Gamma(s_i+\tfrac{1}{2}(\Delta-\tfrac{d}{2}))\left(\tfrac{m_i}2\right)^{-2s_i}e^{i\frac{\pi}2(-s_i+\frac12(\Delta_i-\frac{d}{2}))}\\ \nonumber
&  \times \int_0^\infty\frac{dt_i}{t_i}\frac{t_i^{\Delta_i}}{\Gamma(\Delta_i)}\int_0^\infty\frac{dp_i}{p_i}p_i^{s_i-\frac12(\Delta_i-\frac{d}{2})} \underbrace{\int_{\mathbb{M}_{d+2}} d^{d+2}X\,e^{i(p_1+\ldots +p_n) X^2-2i(t_1 Q_1+\ldots +t_n Q_n)\cdot X}}_{-i\left(\frac{\pi}{-i(p_1+\ldots+p_n)}\right)^{\frac{d+2}2}e^{-\frac{i P^2}{p_1+\ldots+p_n}}}\\ \nonumber
&=\mathcal{N}\pi^{\frac{d+2}{2}}e^{i\frac{\pi}2\frac{d}{2}}
\int[ds_i]_n \prod_i\,\Gamma(s_i+\tfrac{1}{2}(\Delta-\tfrac{d}{2}))\left(\tfrac{m_i}2\right)^{-2s_i}e^{i\frac{\pi}2(-s_i+\frac12(\Delta_i-\frac{d}{2}))}\\ \nonumber & \hspace*{2cm} \times \int_0^\infty\frac{dt_i}{t_i}\frac{t_i^{\Delta_i}}{\Gamma(\Delta_i)}\underbrace{\int_0^\infty\frac{dp_i}{p_i}p_i^{s_i-\frac12(\Delta_i-\frac{d}{2})}(p_1+\ldots+p_n)^{\frac{\Delta_1+\ldots+\Delta_n-d-2}2}}_{\frac{\prod_i\Gamma(s_i-\tfrac{1}{2}(\Delta_i-\tfrac{d}{2}))}{\Gamma\left(\frac{d+2-\Delta_1-\ldots-\Delta_n}2\right)}2\pi i\delta\left(-\tfrac{d+2}2+\sum_j(s_j+\tfrac{d}{4})\right)}\, e^{-i P^2}\\ \nonumber
&=-i\sin\left(\tfrac{\pi}2(\Delta_1+\ldots+\Delta_n-d)\right)\left(\prod_i c^{\text{dS-AdS}}_{\Delta_i}\right) R_{\Delta_1 \ldots \Delta_n}\left(m_1,\ldots,m_n\right)\\ & \hspace*{9.5cm}\times D_{\Delta_1 \ldots \Delta_n}\left(Q_1,\ldots,Q_n\right). \nonumber
\end{align}
where in the first equality we introduced $P=\sum_i t_i Q_i$ and in the last we reabsorbed the normalisation $\mathcal{N}$. $D_{\Delta_1 \ldots \Delta_n}\left(Q_1,\ldots,Q_n\right)$ is the complexified $D$-function \eqref{Danalcont}. The function $R_{\Delta_1 \ldots \Delta_n}\left(m_1,\ldots,m_n\right)$ is given by the Mellin-Barnes integral:
\begin{equation}\label{radialintapp}
   R_{\Delta_1\dots\Delta_n}\left(m_1,\ldots,m_n\right)= \int[ds_i]_n (2\pi i)\delta\left(-\frac{d+2}2+\sum_j\left(s_j+\tfrac{d}{4}\right)\right)\prod_i \widetilde{{\cal K}}^{\left(m_i\right)}_{i\left(\frac{d}{2}-\Delta_i\right)}(s_i)\,,
\end{equation}
where $\widetilde{{\cal K}}^{\left(m_i\right)}_{i\left(\frac{d}{2}-\Delta_i\right)}(s_i)$ is the Mellin transform \eqref{KL-mellin} of the Kontorovich-Lebedev kernel. This can be understood to arise from the integral over the radial direction, which is manifest in the next section (see also \cite{Iacobacci:2022yjo}).

\subsection{In the hyperbolic slicing}
\label{appendix::MINKINTHYPERBOLIC}

We can obtain exactly the same result performing the integral in the hyperbolic slicing of section \ref{sec::HSMINK}, upon which the integral over Minkowski space decomposes into contributions from the regions ${\cal A}_\pm$ and ${\cal D}_\pm$: 
\begin{equation}
    I_{\mathbb{M}_{d+2}} = I_{{\cal A}_+}+I_{{\cal A}_-}+I_{{\cal D}_+}+I_{{\cal D}_-}.
\end{equation}
The integrals in each region can be straightforwardly evaluated along the same lines as the previous section using the parameterisation \eqref{EAdScoords} and \eqref{dScoords} of the regions ${\cal A}_\pm$ and ${\cal D}_\pm$:\footnote{Note that the phases $e^{i\pi(-s_i+\frac{\Delta_i}2-\frac{d}{4})}$ in the first line of contributions $I_{{\cal A}_\pm}$ arise from the $i\epsilon$ prescription $\sqrt{X^2+i\epsilon}$ in the celestial bulk-to-boundary propagator \eqref{explcelesbubo}.} 
\begin{align}
    I_{\mathcal{A}_+}&=\left(\prod_i c^{\text{dS-AdS}}_{\Delta_i}\right)\int_{0}^{\infty}t^{d+1}dt\int[ds_i]_n \prod_i \,\widetilde{ \mathcal{K}}_{i\left(\frac{d}{2}-\Delta_i\right)}^{(m)}(s_i)t^{-d/2-2s_i}e^{i\pi(-s_i+\frac{\Delta_i}2-\frac{d}{4})}\nonumber\\&\hspace*{5cm}\times \prod_i \frac{C^{\text{AdS}}_{\Delta_i}}{i^{\Delta_i}} \int_0^\infty\frac{dt_i}{t_i}\frac{t_i^{\Delta_i}}{\Gamma(\Delta_i)}\int_0^{\infty}\frac{dz}{z^{d+1}}\int d^d\vec{x}\, e^{i\frac{|P|}{z}(1+z^2+x^2)}\,\nonumber \\
    &=\frac{1}2e^{i\pi(-\frac{d}{2}-1+\frac34(\Delta_1+\ldots+\Delta_n))}\left(\prod_i c^{\text{dS-AdS}}_{\Delta_i}\right)R_{\Delta_1 \ldots \Delta_n}\left(m_1,\ldots,m_n\right)\\ & \hspace*{4.5cm}\times \pi^{\frac{d}{2}}\Gamma(\tfrac{\Delta_1+\ldots+\Delta_n-d}2)\prod_i \frac{C^{\text{AdS}}_{\Delta_i}}{i^{\Delta_i}}\int_0^\infty\frac{dt_i}{t_i}\frac{t_i^{\Delta_i}}{\Gamma(\Delta_i)}\,e^{-iP^2}\,. \nonumber
\end{align}
\begin{align}
  \hspace*{-1cm}  I_{\mathcal{A}_-}&=\left(\prod_i c^{\text{dS-AdS}}_{\Delta_i}\right)\int_{-\infty}^{0}(-t)^{d+1}dt \int[ds_i]_n \prod_i \,\widetilde{ \mathcal{K}}_{i\left(\frac{d}{2}-\Delta_i\right)}^{(m)}(s_i)\left(-t\right)^{-d/2-2s_i}e^{i\pi(-s_i+\frac{\Delta_i}2-\frac{d}{4})} \nonumber\\& \hspace*{5cm} \times \prod_i \frac{C^{\text{AdS}}_{\Delta_i}}{i^{\Delta_i}} \int_0^\infty\frac{dt_i}{t_i}\frac{t_i^{\Delta_i}}{\Gamma(\Delta_i)}\int_0^{\infty}\frac{dz}{z^{d+1}}\int d^d\vec{x}\, e^{-i\frac{|P|}{z}(1+z^2+x^2)}\,\nonumber\\
    &=-\frac{1}2e^{i\frac{\pi}2(-d+\frac12(\Delta_1+\ldots+\Delta_n))}\left(\prod_i c^{\text{dS-AdS}}_{\Delta_i}\right)R_{\Delta_1 \ldots \Delta_n}\left(m_1,\ldots,m_n\right) \\ &\hspace*{5cm} \times \pi^{\frac{d}{2}}\Gamma(\tfrac{\Delta_1+\ldots+\Delta_n-d}2)\prod_i \frac{C^{\text{AdS}}_{\Delta_i}}{i^{\Delta_i}}\int_0^\infty\frac{dt_i}{t_i}\frac{t_i^{\Delta_i}}{\Gamma(\Delta_i)}\,e^{iP^2}\,.  \nonumber
\end{align}
\begin{align}
    I_{\mathcal{D}_+}&=\left(\prod_i c^{\text{dS-AdS}}_{\Delta_i}\right)\int_{0}^{\infty}R^{d+1}dR \int[ds_i]_n \prod_i \,\widetilde{ \mathcal{K}}_{i\left(\frac{d}{2}-\Delta_i\right)}^{(m)}(s_i)R^{-d/2-2s_i} \nonumber\\ \nonumber & \hspace*{4cm} \times  \prod_i \frac{C^{\text{AdS}}_{\Delta_i}}{i^{\Delta_i}} \int_0^\infty\frac{dt_i}{t_i}\frac{t_i^{\Delta_i}}{\Gamma(\Delta_i)}\int_{-\infty}^0\frac{d\eta}{(-\eta)^{d+1}}\int d^d\vec{x}\, e^{i\frac{|P|}{(-\eta)}(1-\eta^2+x^2)}\,\\
    &=\frac{1}2e^{i\frac{\pi}2(d-\frac{\Delta_1+\ldots+\Delta_n}2)}\left(\prod_i c^{\text{dS-AdS}}_{\Delta_i}\right)R_{\Delta_1 \ldots \Delta_n}\left(m_1,\ldots,m_n\right)\\ & \hspace*{5cm} \times \pi^{\frac{d}{2}}\Gamma(\tfrac{\Delta_1+\ldots+\Delta_n-d}2)\prod_i \frac{C^{\text{AdS}}_{\Delta_i}}{i^{\Delta_i}}\int_0^\infty\frac{dt_i}{t_i}\frac{t_i^{\Delta_i}}{\Gamma(\Delta_i)}\,e^{-iP^2}\, \nonumber. 
\end{align}
\begin{align} \nonumber 
    I_{\mathcal{D}_-}&=\left(\prod_i c^{\text{dS-AdS}}_{\Delta_i}\right)\int_{0}^{\infty}R^{d+1}dR \int[ds_i]_n \prod_i \,\widetilde{ \mathcal{K}}_{i\left(\frac{d}{2}-\Delta_i\right)}^{(m)}(s_i)R^{-d/2-2s_i} \\& \hspace*{5cm}\times \prod_i \frac{C^{\text{AdS}}_{\Delta_i}}{i^{\Delta_i}} \int_0^\infty\frac{dt_i}{t_i}\frac{t_i^{\Delta_i}}{\Gamma(\Delta_i)} \int_{0}^\infty\frac{d\eta}{\eta^{d+1}}\int d^d\vec{x}\, e^{-i\frac{|P|}{\eta}(1-\eta^2+x^2)}\,\nonumber \\
    &=\frac{1}2e^{-i\frac{\pi}2(d-\frac{\Delta_1+\ldots+\Delta_n}2)}\left(\prod_i c^{\text{dS-AdS}}_{\Delta_i}\right)R_{\Delta_1 \ldots \Delta_n}\left(m_1,\ldots,m_n\right) \\
    & \hspace*{5cm} \times\pi^{\frac{d}{2}}\Gamma(\tfrac{\Delta_1+\ldots+\Delta_n-d}2)\prod_i \frac{C^{\text{AdS}}_{\Delta_i}}{i^{\Delta_i}}\int_0^\infty\frac{dt_i}{t_i}\frac{t_i^{\Delta_i}}{\Gamma(\Delta_i)}\,e^{iP^2}\,. \nonumber 
\end{align}
Notice that the integral over the radial direction in each region - which we placed on the first line of each contribution - is factorised from the integrals in the dS/EAdS directions (the second line). To evaluate the integral over the radial direction to give the function \eqref{radialintapp} we used the identity:
\begin{equation}
    \int^\infty_0 dR\,R^{d+1} R^{-\sum\limits^n_{j=1}\left(2s_j+\frac{d}{2}\right)} = 2\pi i\, \delta\left(-\left(d+2\right)+\sum\limits^n_{j=1}\left(2s_j+\frac{d}{2}\right)\right).
\end{equation}
Interestingly, the contributions from regions ${\cal A}_-$ and ${\cal D}_-$ have opposite $i\epsilon$ prescriptions to the $D$ function defined in \eqref{Danalcont}. These contributions however cancel:
\begin{equation}
    I_{{\cal A}_-}+I_{{\cal D}_-}=0.
\end{equation}
The remaining contributions from regions ${\cal A}_+$ and ${\cal D}_+$ combine to give the result \eqref{IMINK} after reabsorbing the constant ${\cal N}$ given by \eqref{constN}.

\end{appendix}

\newpage

\bibliographystyle{JHEP}
\bibliography{refs}

\end{document}